%% file: EXO-16-012_temp.tex
\pdfoutput=1

\documentclass[11pt,twoside,a4paper,cmspaper,final,collab]{cms-tdr}
\def\svnVersion{435307P}\def\cmsCernNoTag{CERN-EP-2017-027}\def\cmsCernDate{\today}\def\cmsMessage{Published in the Journal of High Energy Physics as \href{http://dx.doi.org/10.1007/JHEP10(2017)180}{\doi{10.1007/JHEP10(2017)180.}}}

\begin{document}\cmsNoteHeader{EXO-16-012}

\hyphenation{had-ron-i-za-tion}
\hyphenation{cal-or-i-me-ter}
\hyphenation{de-vices}
\RCS$HeadURL: svn+ssh://svn.cern.ch/reps/tdr2/notes/EXO-16-XXX/trunk/EXO-16-XXX.tex $
\RCS$Id: EXO-16-XXX.tex 361069 2016-07-29 16:05:25Z khurana $
\newlength\cmsFigWidth
\ifthenelse{\boolean{cms@external}}{\setlength\cmsFigWidth{0.85\columnwidth}}{\setlength\cmsFigWidth{0.4\textwidth}}
\ifthenelse{\boolean{cms@external}}{\providecommand{\cmsLeft}{top\xspace}}{\providecommand{\cmsLeft}{left\xspace}}
\ifthenelse{\boolean{cms@external}}{\providecommand{\cmsRight}{bottom\xspace}}{\providecommand{\cmsRight}{right\xspace}}
\renewcommand{\HGG}{\ensuremath{\Ph\to\gamma\gamma}\xspace}
\newcommand{\Hbb}{\ensuremath{\Ph\to\PQb\PAQb}\xspace}
\newcommand{\mzp}{\ensuremath{m_{\PZpr}}\xspace}
\renewcommand{\Az}{\ensuremath{\cmsSymbolFace{A}}\xspace}
\newcommand{\ptm}{\ensuremath{p_{\mathrm{T}}^{\text{miss}}}\xspace}
\renewcommand{\MET}{\ptm}
\renewcommand{\ETm}{\ptm}
\newcommand{\maz}{\ensuremath{m_{\Az}}\xspace}
\newcommand{\gzp}{\ensuremath{g_{\PZpr}}\xspace}
\newcommand{\MADGRAPHAMC} {\MADGRAPH{}5\_a\MCATNLO}
\newcommand{\x}{\ensuremath{\phantom{0}}}

\cmsNoteHeader{EXO-16-0XX}
\title{Search for associated production of dark matter with a Higgs boson decaying to \bbbar or $\gamma\gamma$ at $\sqrt{s} = 13\TeV$}

\date{\today}

\abstract{A search for dark matter is performed looking for events with large missing transverse momentum
 and a Higgs boson decaying either to a pair of bottom quarks or to a pair of photons. The data
from proton-proton collisions at a center-of-mass energy of 13\TeV, collected in 2015 with the
CMS detector at the LHC, correspond to an integrated luminosity of 2.3\fbinv. Results are
interpreted in the context of a \PZpr-two-Higgs-doublet model, where the gauge symmetry of
the standard model is extended by a U(1)$_{\PZpr}$ group, with a new massive $\PZpr$ gauge boson, and the Higgs sector is extended with four additional Higgs bosons.
In this model, a high-mass resonance $\PZpr$ decays into a pseudoscalar boson A and a light
SM-like scalar Higgs boson, and the A decays to a pair of dark matter particles.
No significant excesses are observed over the background prediction. Combining results from the two
decay channels yields exclusion limits in the signal cross section in the $m_{\PZpr}$- $m_{\textrm{A}}$ phase space.
For example, the observed data exclude the $ \PZpr $ mass range from 600 to 1860\GeV, for $\PZpr$ coupling strength
$g_{\PZpr} = 0.8$, the coupling of A with dark matter particles $g_{\chi}=1$, the ratio of the vacuum expectation values $\tan \beta = 1$,
 and $m_{\mathrm{A}} = 300\GeV$.  The results of this analysis are valid for any dark matter particle mass below 100\GeV. 
}

\hypersetup{%
pdfauthor={CMS Collaboration},%
pdftitle={Search for associated production of dark matter with a Higgs boson decaying to  bb-bar or gamma gamma at sqrt(s) = 13 TeV},%
pdfsubject={CMS},%
pdfkeywords={LHC, CMS, Dark Matter, Mono-Higgs, \texorpdfstring{$E_\textrm{T}^\textrm{miss}$}{MET}, b jet, boosted-jets, 2HDM, physics, software, computing}}

\maketitle

\section{Introduction} \label{sec:intro}

Astrophysical observations have provided strong evidence for the existence of dark matter (DM) in the universe~\cite{FNAL_Review}.
However, its underlying nature remains unknown and cannot be accommodated within the
standard model (SM). The recent discovery of a Higgs boson with mass of about 125\GeV by the ATLAS and
CMS experiments \cite{HiggsObs_ATLAS, HiggsObs_CMS, HiggsObs_CMS_Long}
provides an additional handle to probe the dark sector beyond the SM.
As explained below, in the analyses presented here, it is assumed that there
are five physical Higgs bosons, and that the new state corresponds to the
light neutral CP-even state h.
If DM has origin in particle physics, and if other than gravitational interactions exist between DM and SM particles, DM particles ($\chi$) could be produced at the CERN LHC.
One way to observe DM particles would be through their recoil against a SM particle X (X = g, q, $\gamma$, Z, W, or h)
that is produced in association with the DM.
This associated production of DM and SM particles is often referred to as
mono-X production. The SM particle X can be emitted directly from a quark or
gluon as initial-state
radiation, or through a new interaction between DM and SM
particles, or as final-state radiation.
The Higgs boson radiation from an initial-state quark or gluon is suppressed through Yukawa or loop processes, respectively.
A scenario in which the Higgs boson is part of the
interaction producing the DM particles gives mono-h searches a uniquely enhanced
sensitivity to the structure of couplings between the SM particles and the dark matter~\cite{monoHiggs3,monoHiggs1,2HDM}.
At the LHC, searches for DM in the mono-h channel have been performed
by the ATLAS Collaboration using data  corresponding to integrated
luminosities of 20\fbinv at $\sqrt{s}=8\TeV$ and
3.2\fbinv at $\sqrt{s}=13\TeV$, through the decay channels
\Hbb~\cite{ATLASHBB,ATLAS-2015-PAS} and \HGG~\cite{ATLASHAA}.

\begin{figure}[htbp]
\centering
\includegraphics[width=0.35\textwidth]{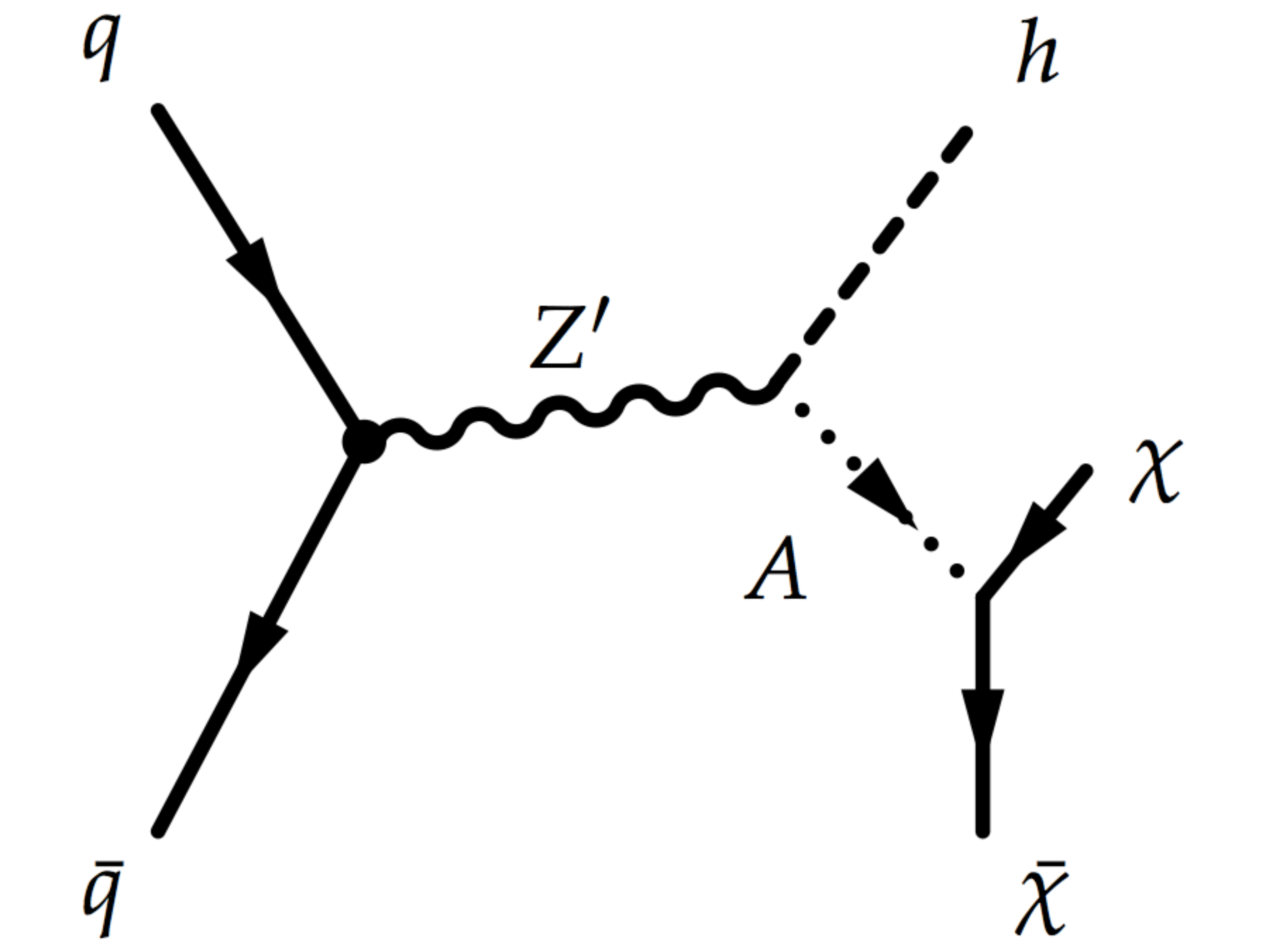}
\caption{Leading order Feynman diagram of the \cPZpr-2HDM ``simplified model''. A pseudoscalar boson \Az decaying into invisible dark matter is produced from the decay of an on-shell $\cPZpr$ resonance. This gives rise to a Higgs boson and missing transverse momentum.}
\label{fig:feynman}
\end{figure}

In this paper, a search for DM is presented in the mono-h channel
in which the Higgs boson decays to either a pair of bottom quarks ($\bbbar$) or photons ($\gamma\gamma$).
The results have been interpreted using a benchmark ``simplified model'' recommended in the ATLAS-CMS Dark Matter Forum, which is described in Ref.~\cite{Abercrombie:2015wmb}: a \cPZpr-two-Higgs-doublet-model (\cPZpr-2HDM)~\cite{2HDM}, where a heavy \cPZpr\ vector boson
is produced resonantly and decays into a SM-like Higgs boson \Ph and an
intermediate heavy pseudoscalar particle \Az, which in turn decays into
a pair of DM particles, as shown in Fig.~\ref{fig:feynman}.

In the \cPZpr-2HDM model, the gauge symmetry of the SM is extended by a
$U(1)_{\cPZpr}$ group, with a new massive \cPZpr\ gauge boson.
A Type-2 2HDM~\cite{Lee:1973iz,Branco:2011iw} is used to formulate
the extended Higgs sector.
A doublet $\Phi_u$ couples only to up-type quarks, and
a doublet $\Phi_d$ couples to down-type
quarks and leptons.
Only $\Phi_u$ and right-handed up-type quarks $u_R$ have an associated charge
under the $U(1)_{\cPZpr}$ group, while $\Phi_d$ and all other SM
fermions are neutral.
After electroweak symmetry breaking, the Higgs doublets attain vacuum
expectation values $v_u$ and $v_d$,
resulting in five physical Higgs bosons:
a light neutral CP-even scalar \Ph, assumed to be the
observed 125\GeV Higgs boson, a heavy neutral CP-even scalar \PH,
a neutral CP-odd scalar \Az, and two charged scalars \Hpm.
The analysis in this paper is performed in the context of the so-called
alignment limit where the \Ph\ has SM-like couplings to fermions and gauge
bosons, and the ratio of the vacuum expectation values $\tan \beta = v_u/v_d > 0.3$, as implied from the perturbativity limit of the Yukawa
coupling~\cite{2HDM,Craig:2013hca} of the top quark, the \Ph-\PH mixing angle $\alpha$ is
related to $\beta$ by $\alpha = \beta - \pi/2$.

The benchmark model is parametrized through six quantities: (i) the pseudoscalar mass
\maz, (ii) the DM mass $m_{\chi}$, (iii) the \cPZpr\ mass \mzp,
(iv) $\tan \beta$, (v) the \cPZpr\ coupling strength \gzp, and (vi)
the coupling constant between the \Az and DM particles $g_{\chi}$.

Only the masses \maz and \mzp affect the kinematic
distributions of the objects in the final states studied in this analysis.
In fact, when \Az is on-shell, i.e. $\maz > 2 m_{\chi}$, the distributions
have little dependence on $m_{\chi}$.
The remaining parameters modify the production cross section of \cPZpr, branching fraction, and decay
widths of the $\cPZpr$ and the \Az, resulting in only small changes to the
final-state kinematic distributions.

This paper considers a \cPZpr\ resonance with mass between 600 and 2500\GeV and
an \Az with mass between 300 and 800\GeV, while the mass of DM particles $m_{\chi}$ is less than or equal to 100\GeV.
The parameters $\tan \beta$ and $g_{\chi}$ are fixed at unity and two different assumptions on \gzp\ are
evaluated as described in more detail later.
Values of \maz below 300\GeV are excluded by constraints on flavor changing
neutral currents from measurements of $\cPqb\to \cPqs\gamma$ \cite{Branco:2011iw},
and are not considered here.

The branching fraction for decays of \Az to DM particles,
$\mathcal{B}(\Az\to\chi\overline{\chi})$, decreases as $m_{\chi}$ increases;
for the range of \maz considered in this paper, the relative decrease of
$\mathcal{B}(\Az\to\chi\overline{\chi})$ is less than 7\% as $m_{\chi}$
increases from 0 to 100\GeV.
Therefore, although signals with $m_{\chi}=100\GeV$ are considered in this search,
the results are valid for any value of dark matter particle mass below 100\GeV.

The results presented here consider only \Az decays to
DM particles and the final signal cross section  $\sigma(\cPZpr \rightarrow \Az \Ph \rightarrow \chi\overline{\chi}\Ph)$
 includes the value of $\mathcal{B}(\Az\to\chi\overline{\chi})$.
With the assumed dark matter particle mass, the value of $\mathcal{B}(\Az\to\chi\overline{\chi})$ is $\approx$ 100\% for $\maz = 300\GeV$.
The branching fraction starts to decrease for $\maz$ greater than twice the mass
of the top quark as the decay $\Az\to$ $\cPqt\cPaqt$ becomes kinematically
accessible. For example, if $\maz = 400\,(800)\GeV$, $\mathcal{B}(\Az\to\chi\overline{\chi})$
reduces to $54\,(42)\%$. 

\begin{figure}[htbp]
\centering
\includegraphics[width=0.45\textwidth]{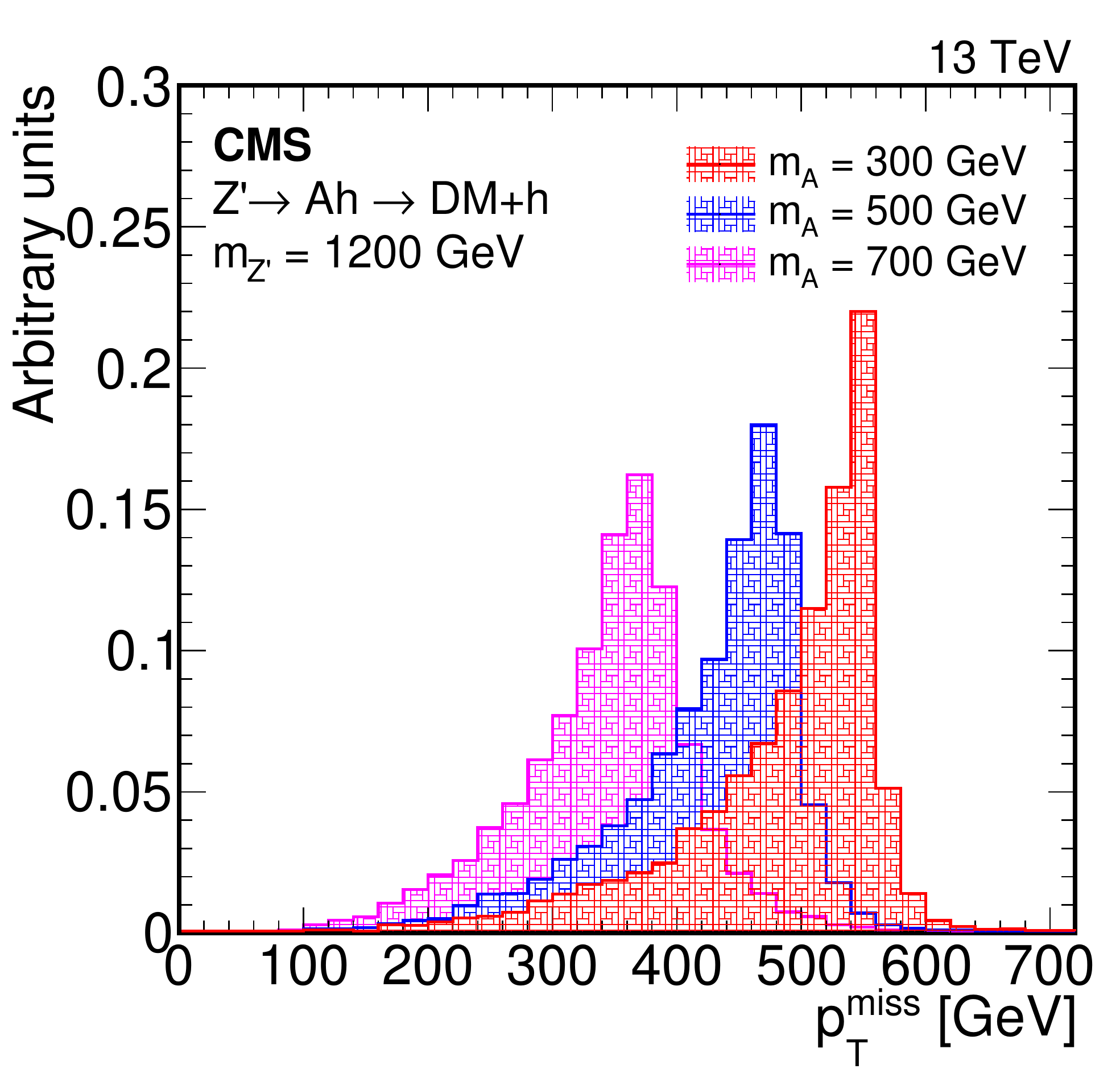}
\caption{Distribution of \MET at generator level for \cPZpr\ $\to$ \Az~h $\to$ DM+h with \maz = 300, 500, and 700\GeV with \mzp = 1200\GeV. All other parameters of the model are fixed, as mentioned in the text.}
\label{fig:mA0}
\end{figure}

The quantity \ptvecmiss, calculated as the negative vectorial sum of the transverse momentum (\pt) of all objects identified in an event, represents the total
momentum carried by the DM particles.
The magnitude of this vector is referred to as \MET.
For a given value of \mzp, the \pt of the \Az decreases as \maz increases.
Therefore, the \MET spectrum softens with increasing \maz.
A comparison of the \MET distributions for three values of \maz is shown in Fig.~\ref{fig:mA0}.

The signal
cross section is calculated
for two assumptions on \gzp:
(i) a fixed value of \gzp = 0.8, as considered in Ref.~\cite{ATLAS-2015-PAS} and recommended in
Ref.~\cite{Abercrombie:2015wmb},
and (ii) using the maximum value from electroweak global fits and
constraints from dijet searches~\cite{2HDM}:
\begin{equation}
\label{eq:gz}
\gzp = 0.03\, \frac{g_{\PW}}{\cos\theta_{\PW}\sin^{2}\beta} \, \frac{\sqrt{m_{\PZpr}^{2}-m_{\cPZ}^{2}}}{m_{\cPZ}},
\end{equation}
yielding $\gzp=0.485$ for $\mzp=1\TeV$, and $\gzp=0.974$ for $\mzp=2\TeV$.
It can be seen from Eq.~\ref{eq:gz} that \gzp = 0.8 is the maximum allowed value of \gzp\ for
$\tan \beta=1$ and $\mzp=1.7\TeV$ (the best reach of LHC as estimated by Ref.~\cite{2HDM}).
Note that this analysis does not consider the contribution of another decay that gives a similar mono-h
signature: $\cPZpr\to \cPZ \Ph$ where $\cPZ\to\nu\overline{\nu}$.
The ratio of branching fractions,
$\mathcal{B}(\cPZpr\to \cPZ \Ph,\cPZ\to\nu\overline{\nu})/\mathcal{B}(\cPZpr \to \Az \Ph,\Az\to\chi\overline{\chi})$,
is a function of $\tan \beta$ and \mzp and does not depend on \gzp since
the value of \gzp cancels in the ratio.

The \Hbb  decay mode has the largest branching fraction (${\approx} 58 \%$) of all, but suffers from relatively poor mass resolution of about 10\%, and while the \HGG branching fraction is small (${\approx} 0.2\%$), the channel benefits from the high precision in reconstructed diphoton mass, with a resolution of about 1--2\%.

\par In the \Hbb channel, the fact that
the \pt of the \Ph\ should increase with \mzp and
decrease with \maz is exploited.
The minimum separation in the pseudorapidity and azimuth ($\eta$, $\phi$) plane
between the decay products of h scales as $m_{\Ph}/p_{\mathrm{T}}^{\Ph}$, where
$p_{\mathrm{T}}^{\Ph}$ is the transverse momentum of the h boson.
The allowed mass ranges of \mzp and \maz
imply a very wide range of values for $p_{\mathrm{T}}^{\Ph}$
and consequently a wide range in the separation of the decay products.
Analysis in this channel is therefore divided into two regimes: (i) a
resolved regime where the \Ph decays to two distinct reconstructed b jets,
and (ii) a Lorentz-boosted regime where the \Ph is reconstructed as a single fat jet.
For each mass point, the analysis with best sensitivity for the expected limit is used as the final result.
The signal extraction is performed through a simultaneous fit to the signal- and background-enriched
control regions.

The search in the \HGG channel is performed by seeking an excess of events over the SM prediction in the diphoton mass spectrum, after requiring a large \MET.
Control samples in data are used to estimate the reducible background, which mainly consists of diphoton SM production. A counting approach is used
to estimate the potential signal.

The paper is organized as follows. After a brief introduction to the
CMS detector in Section~\ref{sec:detector},
the data and simulated events used for the analysis are described in Section~\ref{sec:datasimulation}.
The event reconstruction is detailed in Section~\ref{eventreco}. Section~\ref{sec:eventselection} describes the analysis strategy for both Higgs
boson decay channels.
The description of the most relevant systematic uncertainties affecting the
analysis is found in Section~\ref{sec:systematics}.
Finally, the results of the search are reported in Section~\ref{sec:results},
and the summary is presented in Section~\ref{sec:conclusion}.

\section{The CMS detector}\label{sec:detector}

The central feature of the CMS detector is a superconducting solenoid, of 6\unit{m}
internal diameter, providing an axial magnetic field of 3.8\unit{T} along the beam
direction.
Within the solenoid volume are a silicon pixel and strip
tracker, a lead tungstate crystal electromagnetic calorimeter (ECAL),
and a brass and scintillator hadron calorimeter (HCAL).
Charged particle trajectories are measured by the silicon pixel
and strip tracker system, covering $0 \le \phi \le 2\pi$ in azimuth and
$\abs{\eta}<2.5$.
The electromagnetic calorimeter, which surrounds the tracker volume, consists
of 75,848 lead tungstate crystals
that provide coverage in pseudorapidity $\abs{\eta} < 1.48$ in the barrel region
(EB) and $1.48 < \abs{\eta} < 3.0$ in two endcap regions (EE).
The EB modules are arranged in projective towers.
A preshower detector consisting of two planes of silicon sensors interleaved
with a total of three radiation lengths of lead is located in front of the EE.
In the region $\abs{\eta} < 1.74$, the HCAL cells have widths of $0.087$ in
pseudorapidity and azimuth. In the ($\eta$, $\phi$) plane and for
$\abs{\eta} < 1.48$, the HCAL cells map on to 5${\times}$5 ECAL crystal arrays to form
calorimeter towers projecting radially outwards from the nominal
interaction point. For $\abs{ \eta } > 1.74$, the coverage of the towers increases progressively to a maximum of 0.174 in $\Delta \eta$ and $\Delta \phi$.
Extensive forward calorimetry complements the coverage provided by the barrel
and endcap calorimeters.
Muons are measured in gas-ionization detectors
embedded in the steel return yoke.
A more detailed description of the CMS detector can be found in
Ref.~\cite{Chatrchyan:2008aa}.

\section{Data and simulated samples}\label{sec:datasimulation}

The analysis is performed with pp collision data at
$\sqrt{s} = 13\TeV$ collected by the CMS experiment at the LHC during 2015, corresponding to an integrated
luminosity of 2.3\fbinv.

The \MADGRAPHAMC~v2.3.0~\cite{bib:MADGRAPH}
generator is used to generate the mono-h signal at leading order (LO) as predicted by
the \cPZpr-2HDM model described in
Section~\ref{sec:intro}. In the \MADGRAPHAMC\ generation, a vector particle \cPZpr\ that decays
to a SM-like Higgs boson \Ph\ with mass 125\GeV is produced resonantly together with a heavy pseudoscalar particle A that decays into a pair of DM particles.
The decay of the SM-like Higgs boson is handled by \PYTHIA~8.205~\cite{bib:PYTHIA}.

The associated production of a SM Higgs boson and a Z boson (Z\Ph)
is a small but irreducible background for both decay channels.
The V\Ph\ (Z\Ph\ and W\Ph) processes are
simulated using \POWHEG v2.0~\cite{powheg,Luisoni:2013kna} and
\MADGRAPHAMC for $\qqbar$ and gluon-gluon fusion, respectively.
In the \HGG channel, additional resonant but reducible
backgrounds are considered.
These backgrounds include the SM Higgs boson, produced through
gluon fusion (ggh), through vector boson fusion (VBF),
and in association with top quarks ($\ttbar\Ph$). All of these resonant
backgrounds are modeled at next-to-leading order (NLO) in simulation.
The VBF Higgs boson samples are generated using \POWHEG~\cite{Nason:2009ai},
while the ggh and $\ttbar\Ph$ samples are generated with \MADGRAPHAMC.

The dominant background processes for the \Hbb decay channel
are events with top quarks and W/Z bosons produced in association
with jets. The $\ttbar$ events, produced via the strong interaction,
and electroweak production of single top quarks in the t- and
tW-channels are generated at NLO with
\POWHEG~\cite{Nason:2004rx,Frixione:2007vw,Alioli:2010xd,Frixione:2007nw,Re:2010bp}.
The $s$-channel process of single top quark production
is generated with \MADGRAPHAMC.
Differential measurements of top quark pair production show that the measured
\pt spectrum of top quarks is softer than the one produced in simulation.
Scale factors to correct for this effect are derived from previous CMS
measurements~\cite{TOP-11-013,CMS-PAS-TOP-12-027}.
The sum of top quark pair events and single top quark events is referred to as
``Top quark background'' in the rest of the paper.
The W and Z boson production in association with jets is
simulated at LO with \MADGRAPHAMC.
Up to four additional partons in the matrix element calculations are included.
The MLM matching scheme~\cite{mlm} is used as an interface to the parton shower
generated with \PYTHIA. The cross sections for W+jets and Z+jets processes
are normalized to the next-to-next-to-leading order cross section,
computed using \FEWZ~v3.1~\cite{Li:2012wna}. Moreover, to improve the description of the
distribution of high \pt
W+jets and Z+jets processes, events are reweighted using the generated
\pt of the vector boson to account for NLO quantum chromodynamics (QCD) and electroweak (EW)
contributions~\cite{Kuhn:2005gv,Kallweit:2014xda,Kallweit:2015dum}.
The small background from diboson (WW, WZ, and ZZ) processes, labeled as VV in the rest of the paper, is simulated with \PYTHIA.

For the \HGG decay channel, several nonresonant
background sources can mimic the signal when an event has mismeasured
\ETm and two photons with an invariant mass close to
the mass of the SM-like Higgs boson. These sources include contributions from dijet and
multijet events, EW processes such as
t, $\ttbar$, Z, ZZ, or W bosons produced in association
with one or two photons, $\PGg\PGg$, $\PGg {+} \text{jet}$, and Drell--Yan (DY) production in
association with jets, where the Z boson decays to pairs of electrons and
neutrinos.
These backgrounds are generated with \MADGRAPHAMC, with the exception of the \cPZ\cPZ\ sample,
which is generated with \POWHEG~\cite{Nason:2013ydw}.
These nonresonant background samples are not used for the background estimation, but are used to optimize the selection.

All simulated samples use the NNPDF~3.0 PDF sets~\cite{Ball:2014uwa}.
The parton showering and hadronization are performed with \PYTHIA using
the CUETP8M1 tune~\cite{Skands:2014pea,CMS-PAS-GEN-14-001}.
For the \Hbb\ decay channel, to perform systematic studies in the boosted regime,
an additional signal sample is generated with \MADGRAPHAMC, parton-showered and
hadronized by \HERWIGpp~v2.7.1~\cite{herwigpp} using the UE-EE-5C
tune~\cite{Gieseke:2012ft,Seymour:2013qka}.
The samples are processed through a
\GEANTfour-based~\cite{Agostinelli:2002hh} simulation of the CMS detector.
All samples include the simulation of ``pileup'' arising from additional inelastic
proton-proton interactions in the same or neighboring bunch crossings.
An average of approximately ten pileup interactions per bunch crossing is included in the simulation with a separation between bunches of 25\unit{ns}.
The simulated pileup distribution is reweighted to match the corresponding observed distribution in the analyzed data.

\section{Event reconstruction}
\label{eventreco}
A global event reconstruction is performed using the particle-flow (PF) \cite{CMS-PAS-PFT-09-001, ParticleFlow, Sirunyan:2017ulk} algorithm, which optimally combines the information from all the subdetectors and
produces a list of stable particles, namely muons, electrons, photons,
charged and neutral hadrons.

The reconstructed interaction vertex with the largest value of $\sum_{i} p_{\mathrm{T}i}^{2}$, where $p_{\mathrm{T}i}$ is the transverse momentum of the $i^\mathrm{th}$ track
associated with the vertex, is selected as the primary event vertex.
This vertex is used as the reference vertex for all objects reconstructed using the PF algorithm. The offline selection requires all events to have at least one
primary vertex reconstructed within a 24\unit{cm} window along the z-axis around the mean interaction point, and a transverse distance from the mean interaction region less than 2\unit{cm}.

Jets are reconstructed from the
PF candidates, after removing charged hadrons originating from pileup vertices, using the anti-$\kt$ clustering algorithm~\cite{Cacciari:2008gp} with distance parameters of 0.4 (AK4 jet) and 0.8 (AK8 jet),
as implemented in the \FASTJET package~\cite{bib:fastjet}.
In order to improve the discrimination of signal against multijet background,
the pruning algorithm described in Refs.~\cite{Ellis:2009su,Ellis:2009me}, which is designed to remove contributions from
soft radiation and pileup, is applied to AK8 jets. The pruned jet mass
($m_\text{corrected}^\text{pruned}$)
is defined as the invariant mass associated with the four-momentum of the pruned jet, after the application of
the jet energy corrections~\cite{Khachatryan:2016kdb}.
Corrections to jet momenta are further propagated to the \MET calculation~\cite{CMS-PAS-JME-16-004}.
In addition, tracks with $\pt > 1\GeV$, $\abs{\eta} < 2.5$, and with longitudinal impact parameter $\abs{d_Z}< 0.1\unit{cm}$ from the primary vertex are used to reconstruct the
track-based missing transverse momentum vector, $\vec{p}_{\mathrm{\, T,trk}}^{\text{\, miss}}$.

The jets originating from the decay of b quarks are identified using the
combined secondary vertex (CSV) algorithm \cite{BTV-15-001, BTV-paper-2012-001}, which
uses PF jets as inputs. The algorithm combines the information from the
primary vertex, track impact parameters, and secondary vertices within the jet using a
neural network discriminator. The loose (medium) working point (WP) used in
this analysis has a b jet selection efficiency of 83\% (69\%), a charm jet selection efficiency of 28\% (20\%), and a mistag rate for light-flavor jets of
${\approx}10$\% (1\%)~\cite{BTV-15-001}. The AK8 jets are split into two
subjets using the
soft-drop algorithm~\cite{Dasgupta:2013ihk,softdrop}.
The CSV algorithm is tested and validated for AK4 and
AK8 jets~\cite{BTV-15-001}.
The working points for the analyses of the resolved and boosted regimes were chosen by maximizing the expected significance. The loose WP of the subjet b tagging algorithm is used for the
  boosted regime, whereas the medium WP of the AK4 jet b tagging algorithm is used for the resolved regime, since the background is higher in this case.

Photons are reconstructed in the CMS detector from their energy deposits in the ECAL, which come from an electromagnetic shower involving several crystals. The energy is clustered at the ECAL level by building a cluster of clusters, supercluster (SC), which is extended in the $\phi$ direction because of the strong magnetic field inside the detector, which
deflects the electron and positron produced if the photon converts in the tracker~\cite{tdr_physpreform}.
In order to achieve the best photon energy resolution,
corrections are applied to remove channel-to-channel response variations and to recover energy losses due to incomplete containment of the shower or conversions, as detailed in Ref.~\cite{egamma_phoReco8TeV}.
Additional residual corrections are made to the measured energy scale of the photons in data ($\le$1\%) and to the energy resolution in simulation ($\le$2\%) based on a detailed study of the mass distribution of $\cPZ\to\Pep\Pem$ events.
The uncertainties in the measurements of the photon energy scale and resolution are taken as systematic uncertainties as described in Section~\ref{sec:systematics}.
This process is outlined for the 8\TeV data set in Ref.~\cite{egamma_phoReco8TeV}. Values are adjusted for the 13\TeV data set.

Electron reconstruction requires the matching of a supercluster in the ECAL with a track in the silicon tracker.
Identification criteria~\cite{Khachatryan:2015hwa} based on the ECAL shower shape. 
Muons are reconstructed by combining two complementary algorithms \cite{CMSMuonJINST}:
one in which tracks in the silicon tracker are matched to a muon track segment, and another in which a global track fit is performed, seeded by the muon track segment.
Further identification criteria are imposed on muon candidates to reduce the number of misidentified hadrons. 
Hadronically decaying $\tau$ leptons ($\tauh$) are reconstructed using the hadron-plus-strips
algorithm~\cite{CMSTauJINST}, which uses the charged-hadron and neutral-electromagnetic objects
to reconstruct intermediate resonances into which the $\tau$ lepton decays. 

\section{Event selection and background estimation}\label{sec:eventselection}
\label{eventselection}

This analysis searches for excesses over the background-only prediction in events with large \MET and a system of two b-tagged jets or two photons that has a reconstructed invariant mass close to the mass of the SM-like Higgs boson \Ph.
In the \Hbb decay channel, the analysis relies on fitting the \MET distribution simultaneously in the signal region (SR), defined after selecting a mass window around the Higgs boson mass, and in background-enriched control regions (CRs).
For the \HGG decay channel, a simple analysis is performed where the signal and resonant background contributions are estimated by counting
the number of simulated events in the SR, while the nonresonant background is
extrapolated from the data in a low-\MET region.
In the following sections, the event selection and analysis strategy are described in detail for the two channels separately.

\subsection{\texorpdfstring{The channel \Hbb}{h to bb}}
A search for DM produced in association with \Hbb is performed in a resolved regime, where events are required to have at least two AK4 jets, and in the
Lorentz-boosted regime where one AK8 jet is required.
In addition, \MET is required to be large because it is a key signature of the signal events
and it provides strong rejection against the large reducible backgrounds described in Section~\ref{sec:datasimulation}.

\subsubsection{Event selection}
The trigger used in the selection of signal-like events requires $\MET > 90\GeV$ and $H_{\mathrm{T}}^{\text{miss}} > 90\GeV$, where $H_{\mathrm{T}}^{\text{miss}}$ is defined as the magnitude of the vectorial sum of the \pt of all jets in the event with $p_{\mathrm{T}} > 20\GeV$.
An additional trigger with a $\MET > 170\GeV$ requirement is used to achieve higher efficiency.
In this way, events with either high \MET or high $H_{\mathrm{T}}^{\text{miss}}$ will pass the trigger.
For events passing the selection criteria that have $ \MET > 170 (200)\GeV$ for the resolved (boosted) analysis, the trigger efficiency is found to be greater than 98\%.
The \MET threshold for the analysis of the resolved regime is set slightly lower to enhance the signal efficiency in this region of phase space, where the \MET distribution is softer.

Event filters are used to remove spurious high \MET events caused by instrumental noise in the calorimeters, or beam halo muons. It has been verified that the efficiency of these filters for accepting signal events is very close to 100\%.
The main part of the event selection consists of Higgs boson tagging. This selection is different for the resolved and boosted analyses.
In the resolved regime, events are
required to have two AK4 jets with $\pt>30\GeV$ and $\abs{\eta}< 2.4$.
These two jets are used to reconstruct the Higgs boson candidate, which
is required to have $\pt > 150\GeV$.
Each of the two AK4 jets in the resolved regime is required to pass the b tagging selection, whereas in the boosted regime, the two subjets inside an AK8 jet must both pass the b tagging selection.
In the boosted regime, the decay products from the Higgs boson are merged. Therefore, an AK8 jet with \pt greater than 200\GeV is used to reconstruct
the Higgs boson. If more than one Higgs boson candidate is reconstructed, the ambiguity is resolved by selecting the candidate with the highest \pt.
Backgrounds due to hadronic jets are further reduced by constraining the reconstructed Higgs boson candidate mass, $m_\mathrm{bb}$, to be between 100 and 150\GeV.
For the resolved regime, the Higgs boson candidate mass is reconstructed using two b-tagged AK4 jets.
For the boosted regime, the corrected pruned mass of the AK8 jet with two b-tagged subjets is used as the Higgs boson candidate mass.

Multijet events can act as a source of background  when the energy of one of the jets is mismeasured.
Therefore, the absolute difference between the azimuthal angles of the vector
\ptvecmiss\ and any other AK4 jet with $\pt > 30\GeV$
is required to be greater than 0.4 radians.
Multijet background is further reduced in the
resolved analysis by requiring the azimuthal angle difference between the \ptvecmiss and $\vec{p}_{\mathrm{\, T,trk}}^{\text{\, miss}}$ to be less than 0.7 radians.

Events are rejected if they have any
isolated electron (muon) with $\pt > 10\GeV$ and $\abs{\eta} < $ 2.5 (2.4) or any $\tauh$ candidates with $\pt > 20\GeV$ and $\abs{\eta} < 2.3$~\cite{Khachatryan:2015hwa,Chatrchyan:2013sba,CMSTauJINST}.
In addition, the events must not have any additional loose AK4 b-tagged jet or
more than one additional AK4 jet with $\pt > 30\GeV$ and $\abs{\eta} < 4.5$.
These vetoes considerably reduce the background from semileptonic top decay
modes and leptonic decays of W+jets.

The product of the detector acceptance and selection efficiency varies from
1 to 29\%, depending on the values of
\mzp and \maz. The average \MET increases with \mzp and decreases with \maz.
The overall selection efficiency, shown in Table \ref{tab:AcceptanceHGG},
follows the same trend.

\subsubsection{Analysis strategy and background estimation}

Several CRs are used to
correct the background normalizations with dedicated scale factors. For both resolved and boosted regimes, the selection criteria of these CRs are kept as close as possible to those of the SR, except for the inversion of the additional
object vetoes (leptons, jets) and the Higgs boson mass window.
This makes the CRs orthogonal to the SR.

For the resolved regime, three CRs are specified: Z($\to\nu\overline{\nu}$)+jets,
top quark, and W+jets.
The b tagging selection in all the CRs is the same as in the SR in order to minimize the b tagging systematic uncertainties
when extrapolating the background scale factors measured in the CRs to the SR.
The Z($\to\nu\overline{\nu}$)+jets CR is defined with the same selection as the
SR, except for the inversion of the reconstructed Higgs boson mass requirement.
The W+jets and top quark CRs are defined by removing the mass selection and
requiring exactly one isolated electron (muon) with $\pt > 10\GeV$ and
$\abs{\eta}< 2.5$ (2.4).
Events with one additional AK4 jet are placed in the top quark CR, whereas
events with no additional AK4 jets enter the W+jets CR.

For the boosted regime, the Z($\to\nu\overline{\nu}$)+jets CR is defined by inverting the mass requirement for the AK8 jet.
Owing to the low event count and very similar topology between the W+jets and top
quark backgrounds it is difficult to
construct two separate CRs for W+jets and top quark backgrounds.
Hence, the single-lepton CR, a combination of mainly W+jets and top quark
events, is defined using the same selection as that for the signal, but
requiring exactly one isolated electron (muon) with $\pt > 10\GeV$ and
$\abs{\eta}< 2.5$ (2.4)
and removing the mass requirement.

\begin{figure}[htbp]
\centering
\includegraphics[width=0.45\textwidth]{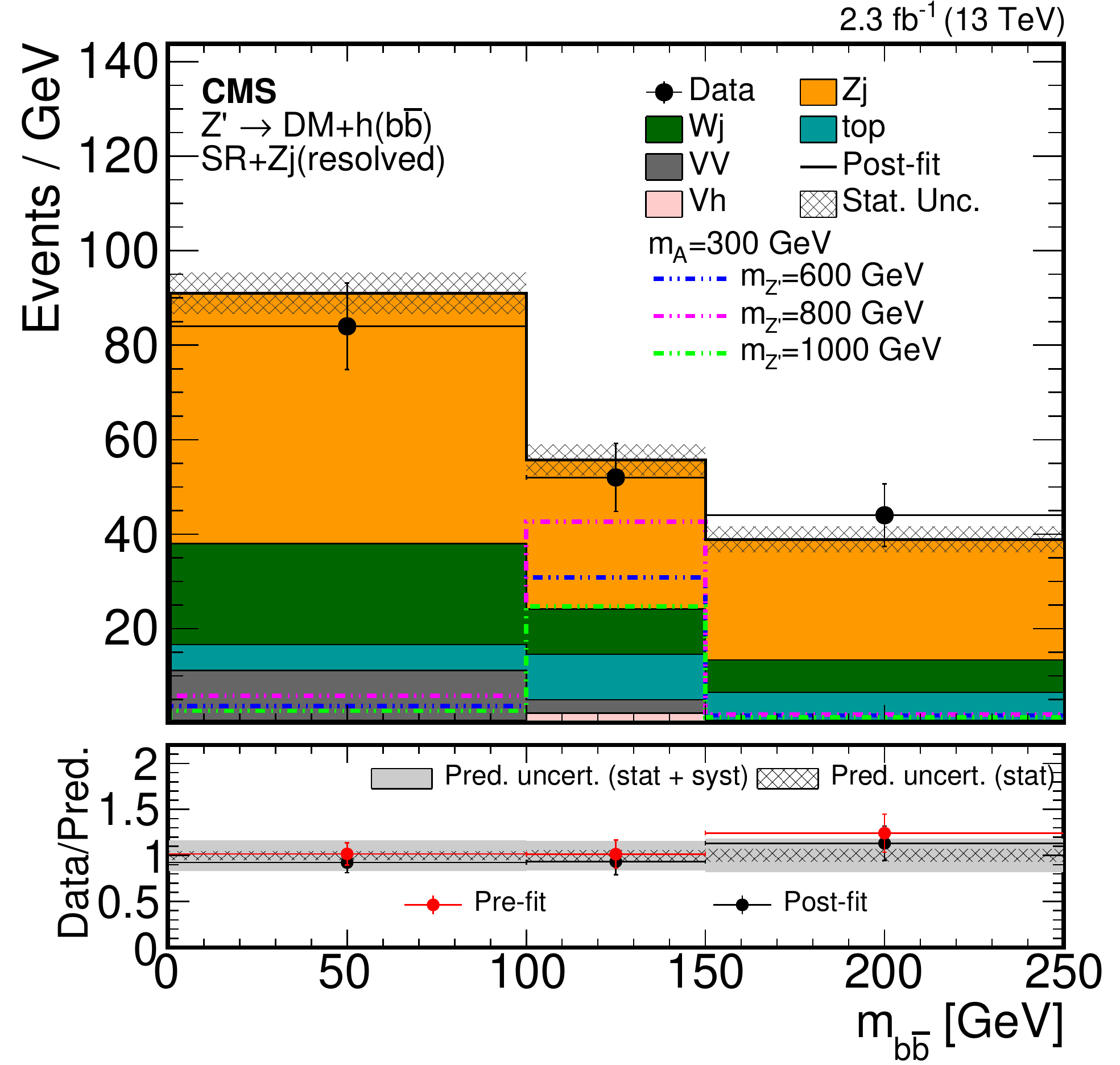}
\includegraphics[width=0.45\textwidth]{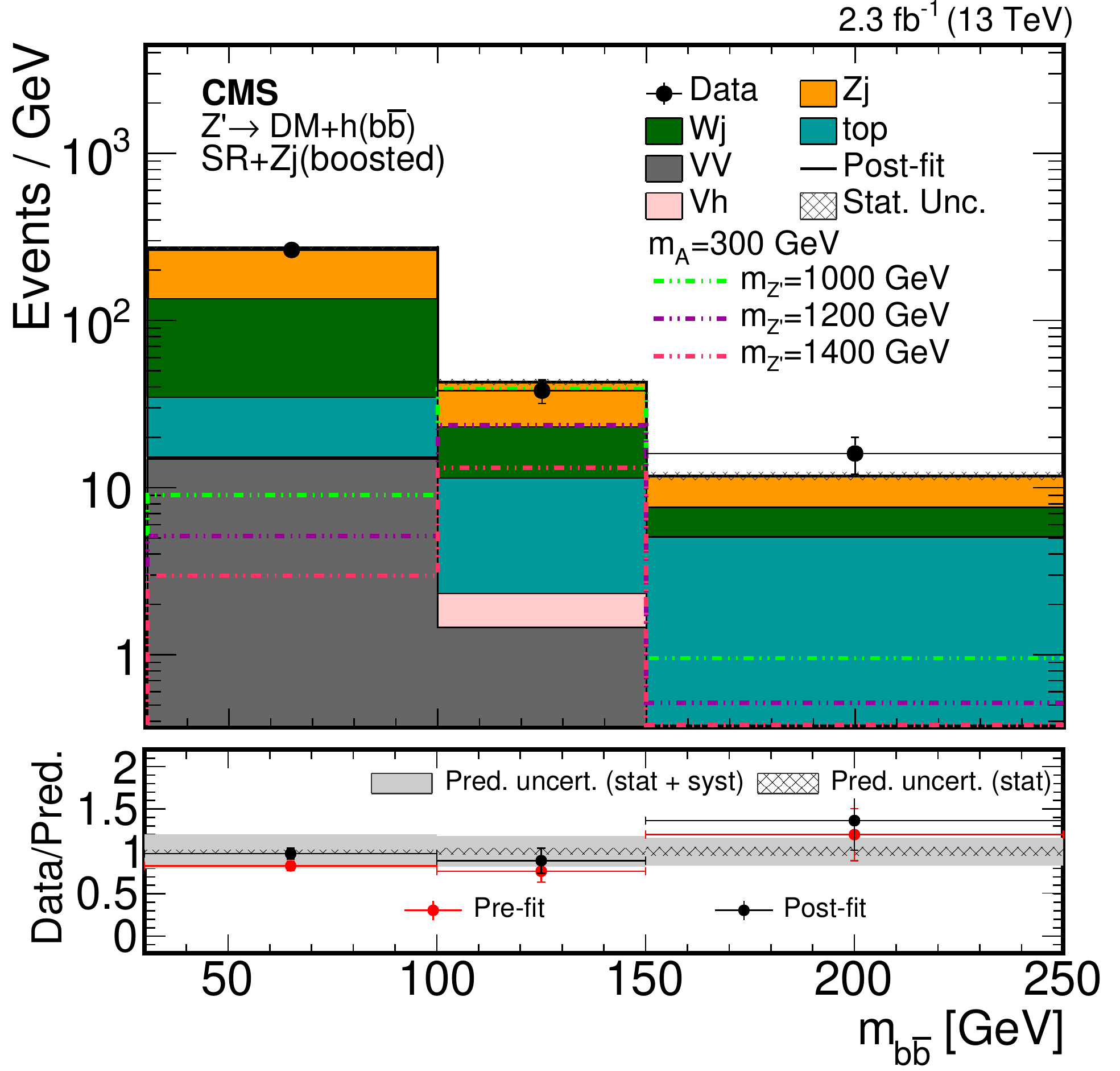}
\caption{Post-fit distribution of the reconstructed Higgs boson candidate mass expected from SM backgrounds and observed in data for the resolved (left)
and the boosted (right) regimes with three different \mzp signal points overlaid. Other parameters for this model are fixed to $m_{\chi} = 100\GeV$ and $\tan{\beta} = g_{\chi} = 1$. The cross sections
for the signal models are computed assuming $\gzp = 0.8$. The bottom panels show the data-to-simulation ratios for pre-fit (red markers) and post-fit (black markers) background predictions with a hatched band corresponding to the uncertainty due to the finite size of simulated samples and a gray band that represents the systematic uncertainty in the post-fit background prediction (see Section~\ref{sec:systematics}). The second bin represents the SR, while the events in the first and third bins are merged and represent the mass sidebands (Z($\to\nu\overline{\nu}$)+jets) CR. }

\label{fig:massHbb}
\end{figure}

\begin{figure}[htbp]
\centering
\includegraphics[width=0.49\textwidth]{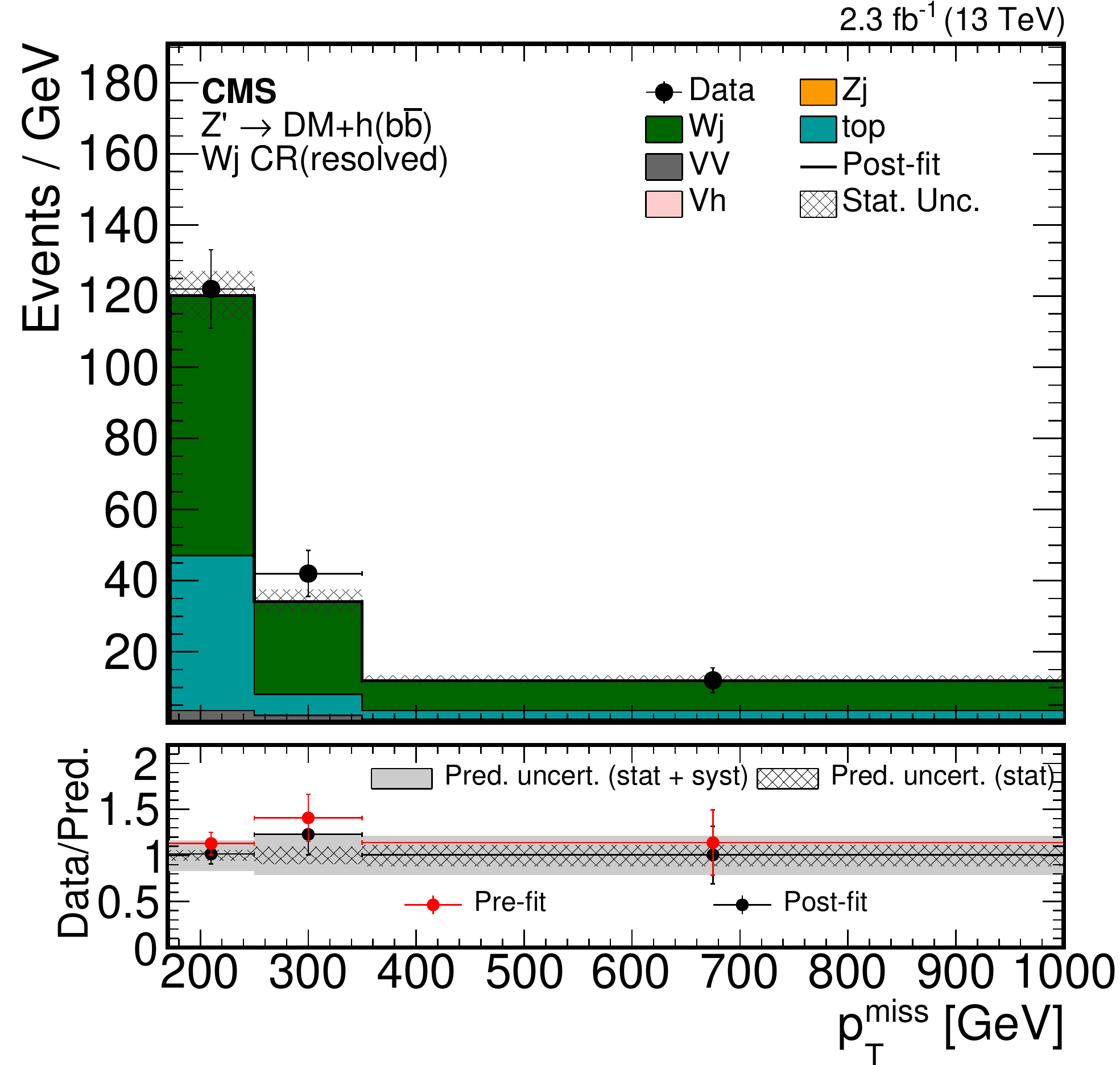}
\includegraphics[width=0.49\textwidth]{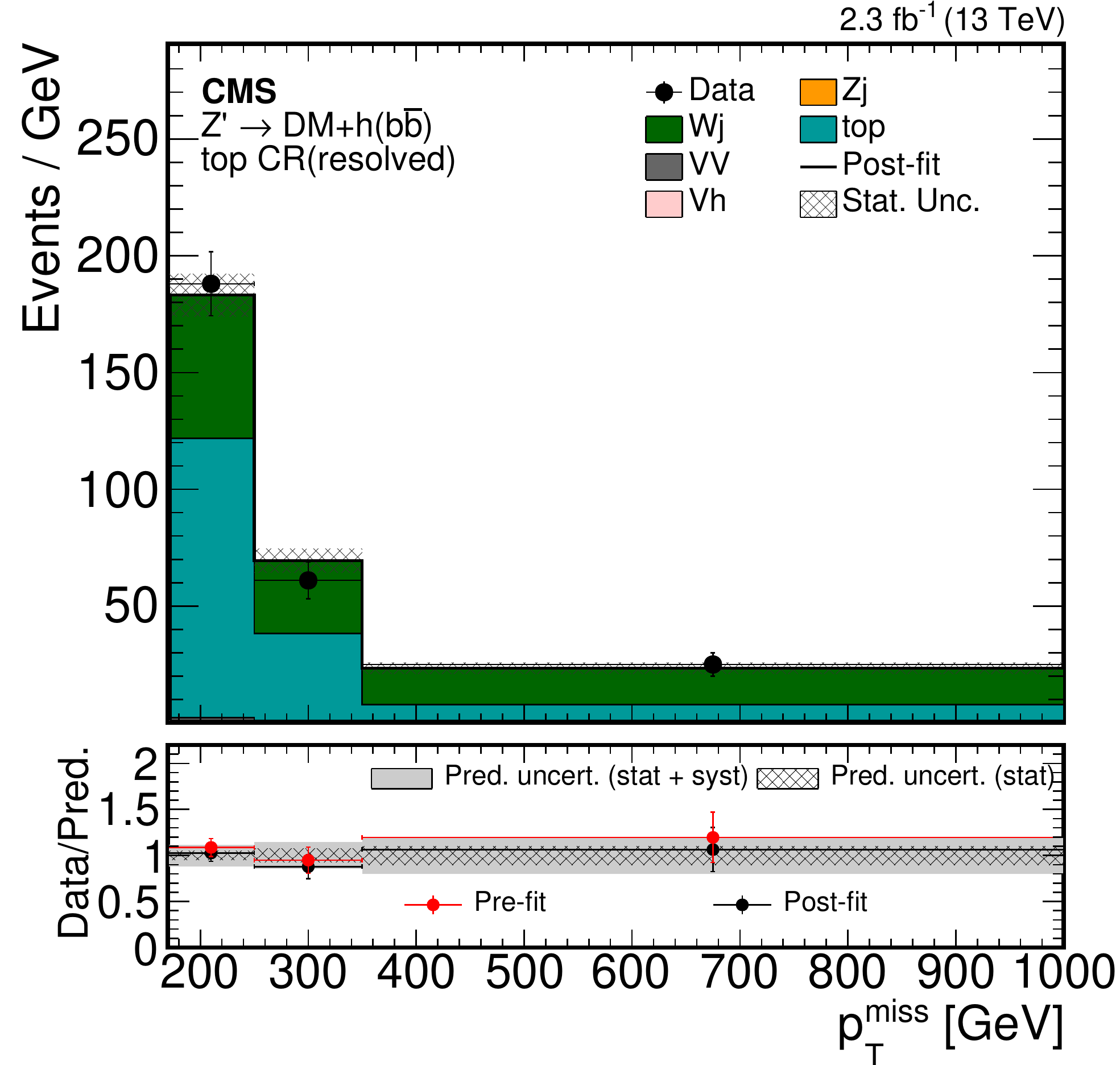}
\includegraphics[width=0.49\textwidth]{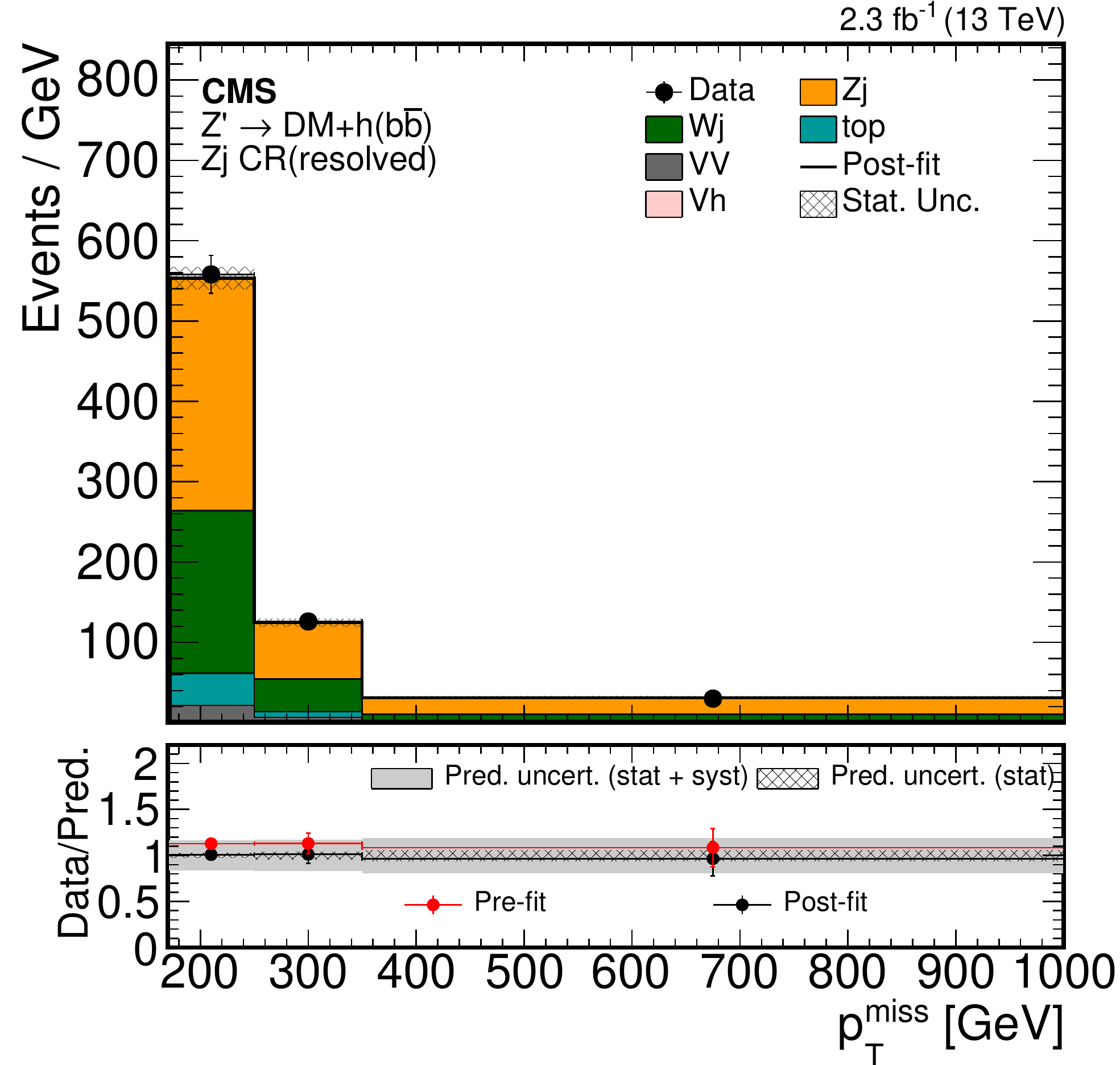}

\caption{Post-fit distribution of \MET expected from SM backgrounds and observed in data for the W+jets (upper left), top quark (upper right) and Z($\to\nu\overline{\nu}$)+jets (lower) CRs for the resolved regime.  The bottom panels show the data-to-simulation ratios for pre-fit (red markers) and post-fit (black markers) background predictions with a hatched band corresponding to the uncertainty due to the finite size of simulated samples and a gray band that represents the systematic uncertainty in the post-fit background prediction (see Section~\ref{sec:systematics}). The last bin includes all events with $\MET > 350\GeV$.}
\label{fig:controlregionR}
\end{figure}

\begin{figure}[htbp]
\centering
\includegraphics[width=0.49\textwidth]{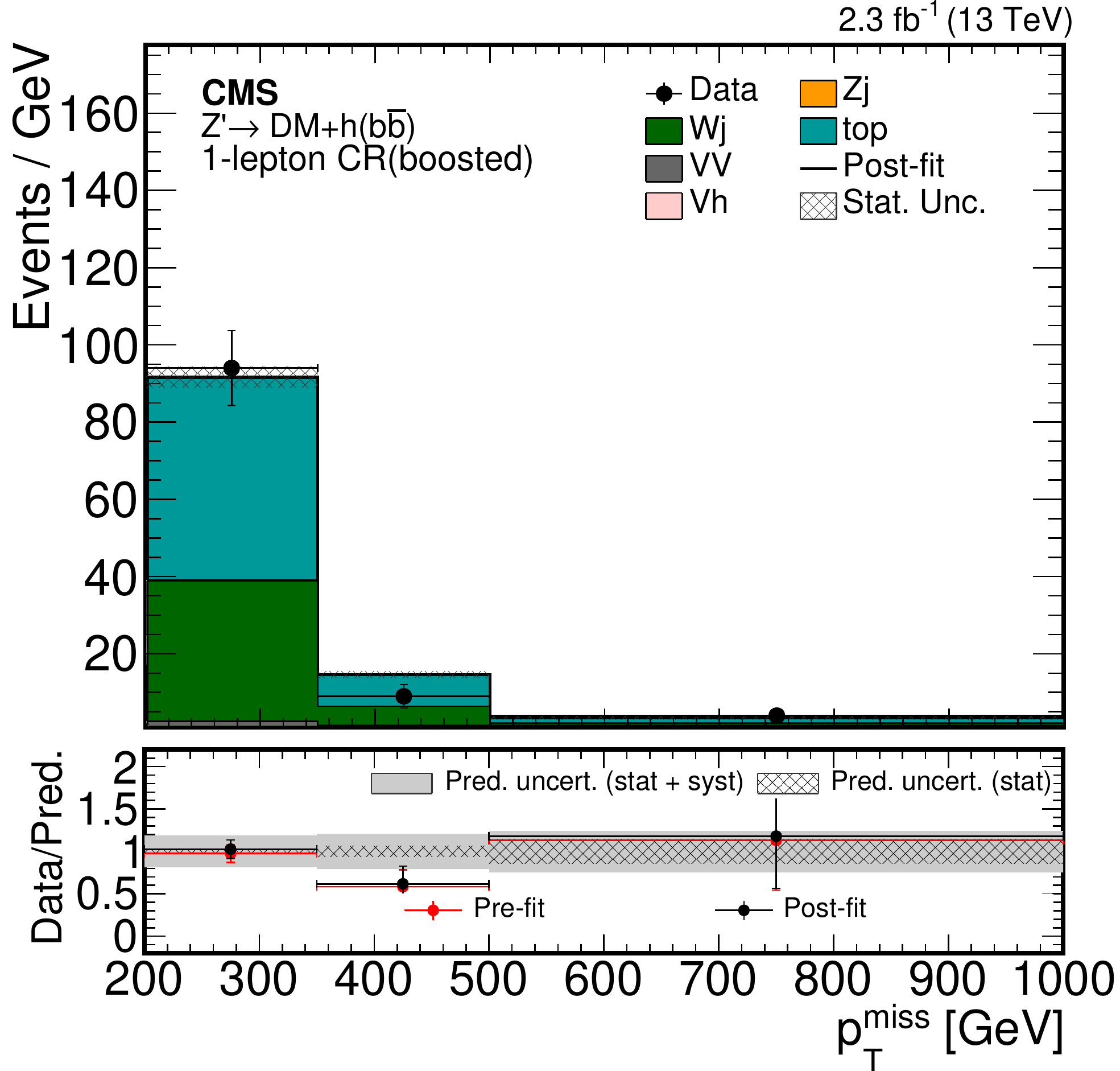}
\includegraphics[width=0.49\textwidth]{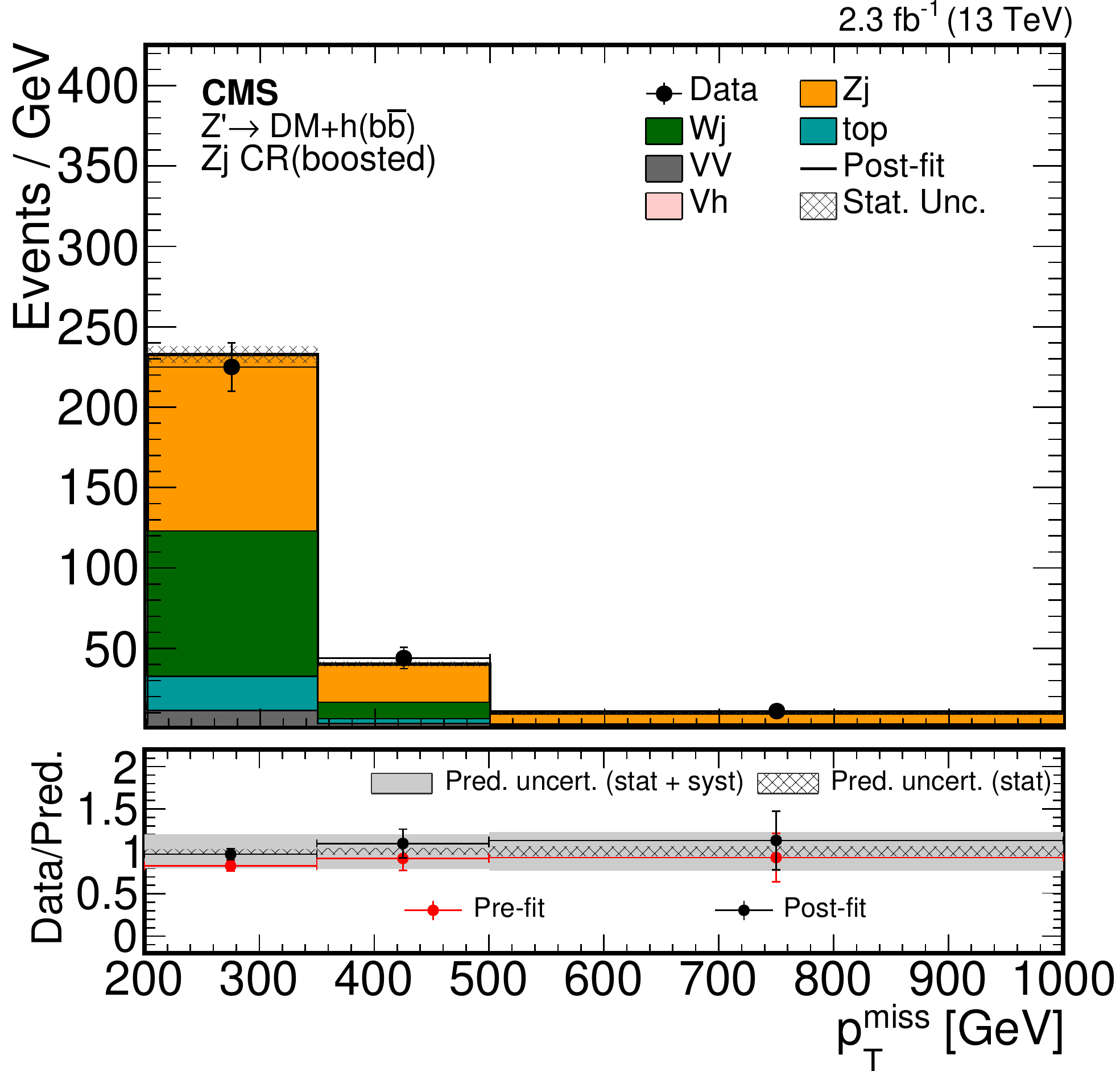}
\caption{Post-fit distribution of \MET expected from SM backgrounds and observed in data for the single-lepton CR and Z($\to\nu\overline{\nu}$)+jets CRs for the boosted regime. The bottom panels show the data-to-simulation ratios for pre-fit (red markers) and post-fit (black markers) background predictions with a hatched band corresponding to the uncertainty due to the finite size of simulated samples and a gray band that represents the systematic uncertainty in the post-fit background prediction (see Section~\ref{sec:systematics}). The last bin includes all events with $\MET > 500\GeV$.}
\label{fig:controlregionB}
\end{figure}

\begin{figure}[htbp]
\centering
\includegraphics[width=0.49\textwidth]{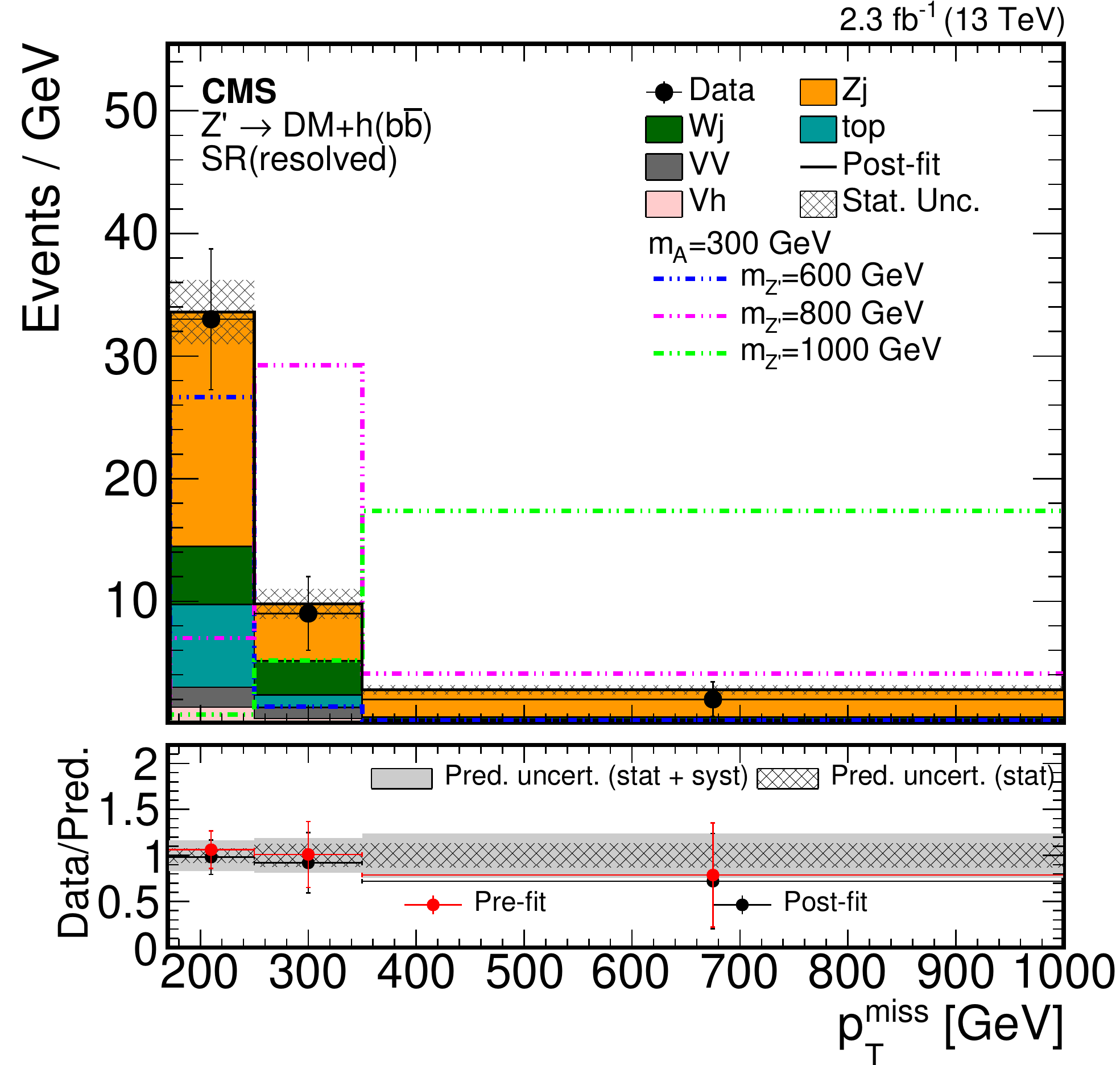}
\includegraphics[width=0.49\textwidth]{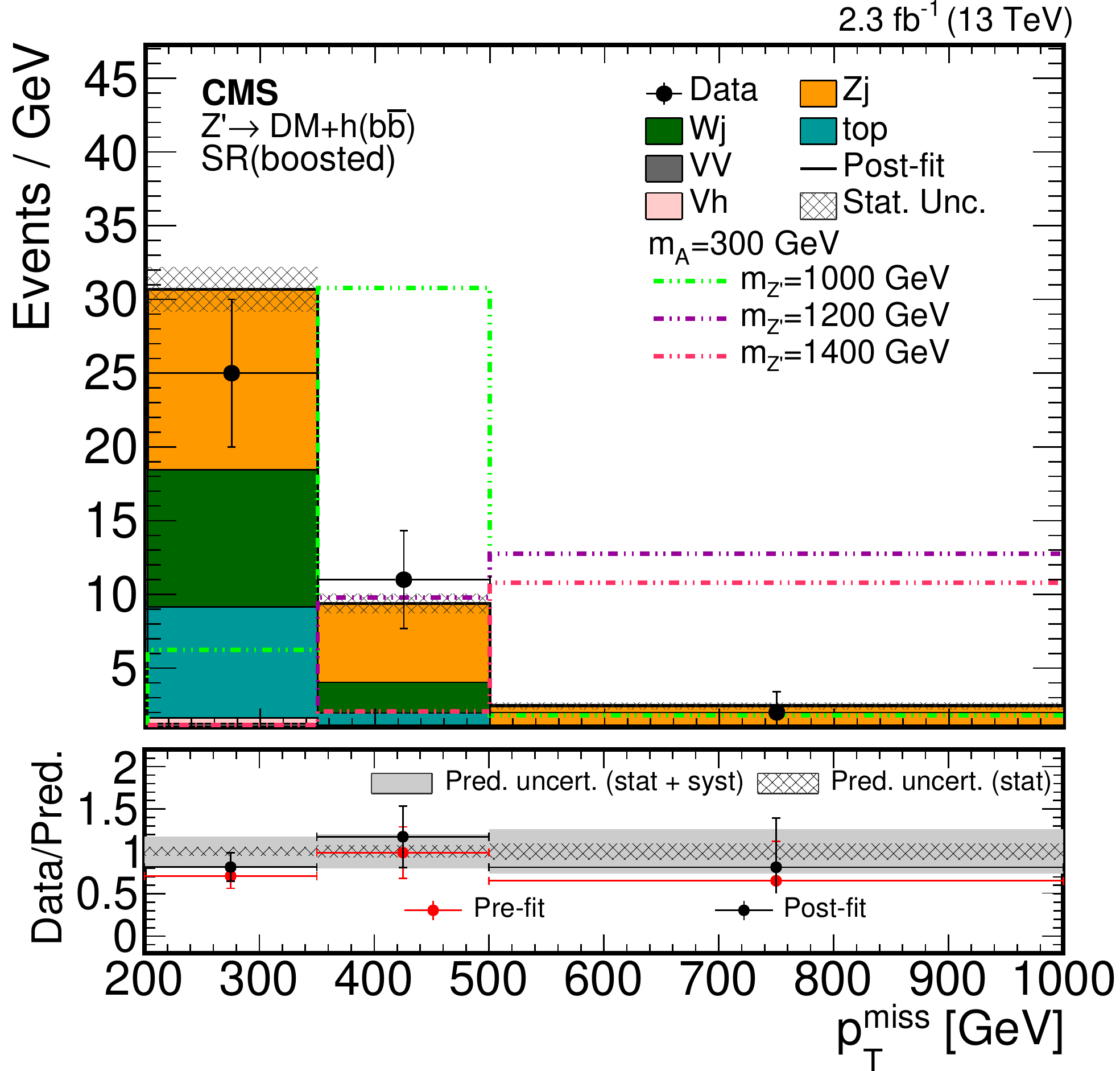}
\caption{Post-fit distribution of \MET expected from SM backgrounds and observed in data for the resolved (left) and the boosted (right) regimes in the signal region with three different \mzp signal points overlaid. Other parameters for this model are fixed to $m_{\chi} = 100\GeV$ and $\tan{\beta} = g_{\chi} = 1$. The cross sections 
for the signal models are computed assuming \gzp = 0.8. The bottom panels show the data-to-simulation ratios for pre-fit (red markers) and post-fit (black markers) background predictions with a hatched band corresponding to the uncertainty due to the finite size of simulated samples and a gray band that represents the systematic uncertainty in the post-fit background prediction (see Section~\ref{sec:systematics}). The last bin includes all events with $\MET > 350 $ (500)\GeV for the resolved (boosted) regime.}
\label{fig:finalSignalPlots}
\end{figure}

Figure~\ref{fig:massHbb} shows the Higgs boson candidate mass for the resolved and boosted regimes.  They correspond to the
simultaneous fit of the \MET distributions in the SR and background enriched CRs to extract the signal.
Data-to-simulation ratios for pre-fit and post-fit background predictions are shown in the lower panels of all Figs.~\ref{fig:massHbb}--\ref{fig:finalSignalPlots}.

Figure~\ref{fig:controlregionR} shows the comparison of data and simulation for the main observable, \MET, in the W+jets, top quark, and Z($\to\nu\overline{\nu}$)+jets  CRs for the resolved regime.
The comparison between data and simulated samples for the boosted regime is shown in Fig.~\ref{fig:controlregionB} for the single-lepton CR and the Z($\to\nu\overline{\nu}$)
mass sideband region.

Figure~\ref{fig:finalSignalPlots} shows the \MET distributions in three bins in the SR that are used for the final signal extraction.
These three bins were chosen to optimize the expected limits.
The selected signal and background events are compared to data and fit simultaneously in the SR and CRs in three \MET bins, separately
for the resolved and the boosted regimes.

The simultaneous fit of SR and background-enhanced CRs is performed correlating the scale factors and systematic uncertainties as described in Section \ref{sec:systematics}.
The measured data-to-simulation post-fit scale factors are compatible with unity within the total combined statistical and systematic uncertainty. In particular, for the resolved regime, the scale factors for the backgrounds are 1.23 $\pm$ 0.17 for Z($\to\nu\overline{\nu}$)+jets,  1.33 $\pm$ 0.19 for W+jets, and 1.13 $\pm$ 0.17 for the top quark
contributions.
For the boosted analysis, the scale factors are 0.77 $\pm$ 0.15 for Z($\to\nu\overline{\nu}$)+jets and 0.95 $\pm$ 0.19 for W+jets and top quark processes.
Although the background scale factors do not show a common trend between the boosted and resolved analyses, it should be noted that the b-tagging requirement, selected phase space and other parameters are different in the two cases. Thus the two simultaneous fits are essentially independent, allowing the post-fit scale factors to move in either direction from unity.

\subsection{\texorpdfstring{The channel \HGG}{h to gg}} \label{sec:photon_id}
The \HGG search is performed using a diphoton selection.
A set of requirements is applied to ensure good-quality photon candidates.
Additional kinematic requirements on the objects in the final state are applied to reduce the background.
The diphoton invariant mass and \MET are used as the discriminating variables to estimate the signal.

\subsubsection{Event selection}
Diphoton triggers with asymmetric transverse energy thresholds (30/18\GeV) are used to select events with the diphoton invariant mass above 95\GeV.
The trigger selection uses a very loose photon identification based on
the cluster shower shape and loose isolation requirements (both defined in detail in Ref.~\cite{egamma_phoReco8TeV}), and a
requirement that the ratio of hadronic-to-electromagnetic energy of the photon candidates is less than 0.1.

The main source of background for photons, which arises from jets with high electromagnetic energy content, is rejected by considering the ratio of energies
deposited by the photon candidate in the hadron and electromagnetic calorimeters and the spread of the energy deposition in the $\eta$ direction, as described
in ~\cite{egamma_phoReco8TeV}. In addition, misidentified photons are rejected using the isolation
variables
$\mathrm{Iso_{Ch}}$, $\mathrm{Iso}_\gamma$, and $\mathrm{Iso_{Neu}}$ calculated by summing the
\PT of the charged hadrons, photons and neutral hadrons, respectively,
in a cone of radius $\DR = 0.3$.
In the photon identification,  $\mathrm{Iso_{Neu}}$ and $\mathrm{Iso}_\gamma$ are corrected for the median transverse energy density ($\rho$) of the event
to mitigate the effects of pileup \cite{pileupSubtract}.

The photons in the EB (\ie the photons with $\left|\eta\right| \leq 1.44$) and photons in the EE
($1.566 \leq \left|\eta\right| \leq 2.5 $) have different selection criteria, equivalent to those used
in Refs.~\cite{CMSHiggsGG1,CMSHiggsGG2}.
The working point chosen for this analysis corresponds to 90.4\% (90.0\%) photon ID efficiency
in the EB (EE),
while the misidentification rate in the EB (EE) is 16.2\% (18.7\%) for objects with $\pt>20\GeV$.

A high-quality interaction vertex, defined as the reconstructed vertex with the largest number of charged tracks, is associated to the two photons in the event.
The efficiency of selecting the correct vertex for all generated mass points, defined as the fraction of signal events with well reconstructed vertices that have a
\emph{z} position within 1\unit{cm} of the generator-level vertex, is approximately 78\%.

The optimal signal selection is chosen by studying the discriminating power of variables such as the $\PT/m_{\gamma\gamma}$ of each photon, \MET, and the \PT of the diphoton system ($p_{\mathrm{T}\gamma\gamma}$).
A selection on \PT that scales with $m_{\gamma\gamma}$ is chosen such that it does not distort the $m_{\gamma\gamma}$ spectrum shape.
The $p_{\mathrm{T}\gamma\gamma}$ variable, included because it has a better resolution than \MET, has a distribution of values that are on average larger for signal than
for background events, given that the Higgs boson is expected to be back-to-back in the transverse plane with the \ptvecmiss.

In addition, two geometrical requirements are applied to enhance the signal over background discrimination and to veto background events with mismeasured \MET:
\begin{itemize}
\item the azimuthal separation between the \ptvecmiss and the
Higgs boson direction (reconstructed from the two photons) $\abs{\Delta\phi (\gamma\gamma, \ptvecmiss)}$ must be greater than 2.1 radians.
\item the minimum azimuthal angle difference between the \ptvecmiss and the jet direction in the event $min(|\Delta\phi (\text{jet}, \ptvecmiss)|)$ must be greater than 0.5 radians.
The jet direction is derived by considering all the jets reconstructed from the clustering of PF candidates by means of the anti-$k_t$ algorithm~\cite{Cacciari:2008gp} with a distance parameter of 0.4. Jets are considered if they have a \pt above 50\GeV  in the $\abs{\eta}$ range below 4.7 and satisfy a loose set of identification criteria designed to reject spurious detector and reconstruction effects.
\end{itemize}

The set of selection criteria that maximizes the expected significance for each $\PZpr$ mass point is studied. The optimized selection for the
$\mzp = 600\GeV$ and $\maz = 300\GeV$ sample maintains a large efficiency for the other signal mass points,
while the backgrounds remain small. Therefore a common set of criteria
is used for all signal masses with \mzp between 600 and 2500\GeV and \maz between 300 and 800\GeV.
The chosen kinematic selections include $p_{\mathrm{T1}}/m_{\gamma\gamma} > 0.5 $, $p_{\mathrm{T2}}/m_{\gamma\gamma} > 0.25$, $p_{\mathrm{T}\gamma\gamma} > 90\GeV$, $\MET > 105\GeV$.
Events are vetoed if they have any muons or more than one electron present.
This allows the analysis to be sensitive to events where an electron originating from conversion of the photon before reaching the ECAL is identified outside the photon supercluster.
Standard lepton identification requirements are used~\cite{Khachatryan:2015hwa,Chatrchyan:2013sba}.
This requirement is 100\% efficient for the signal and reduces significantly the EW background contributions.

\begin{figure}[h]
\centering
\includegraphics[width=0.45\textwidth]{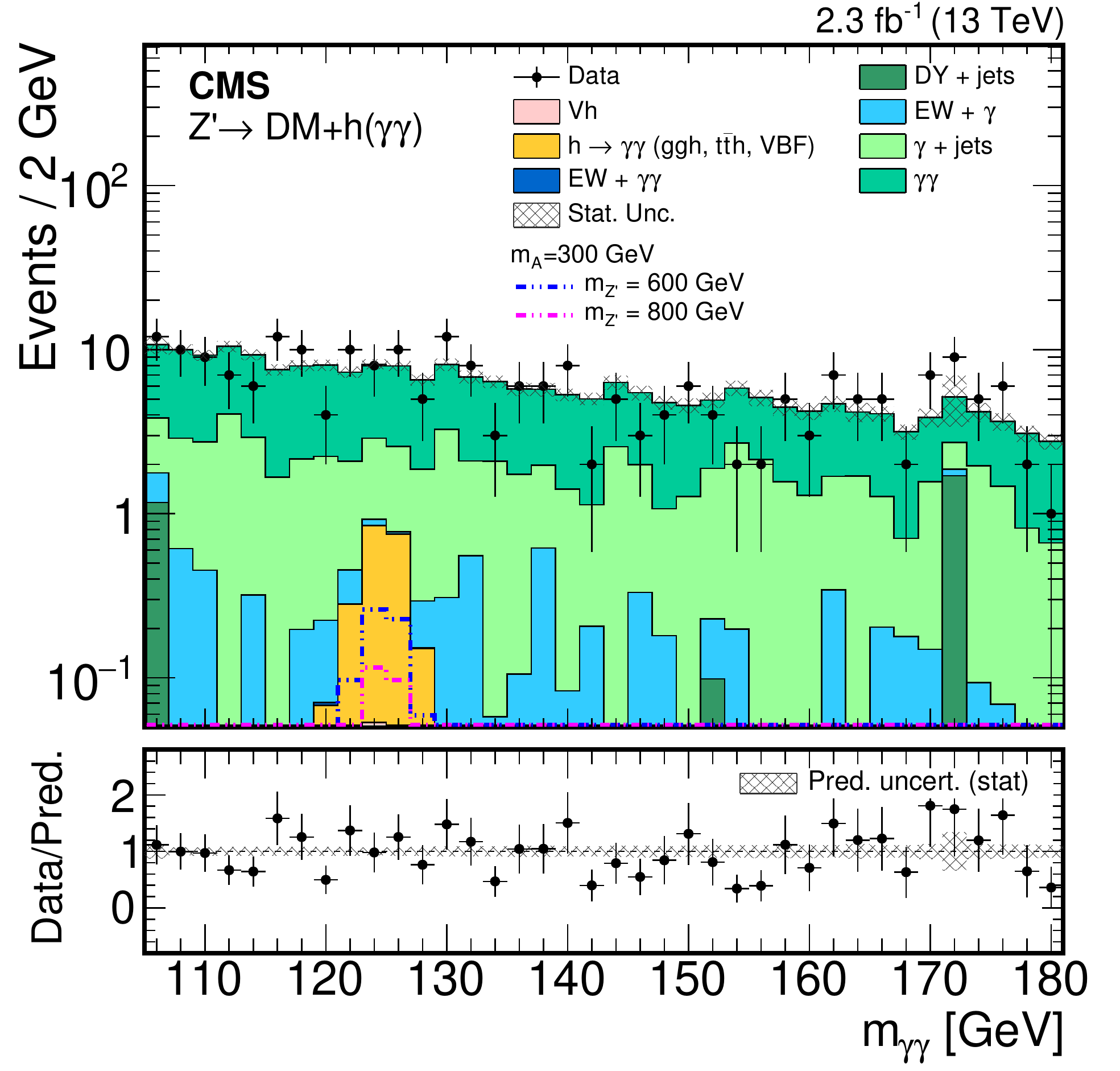}
\includegraphics[width=0.45\textwidth]{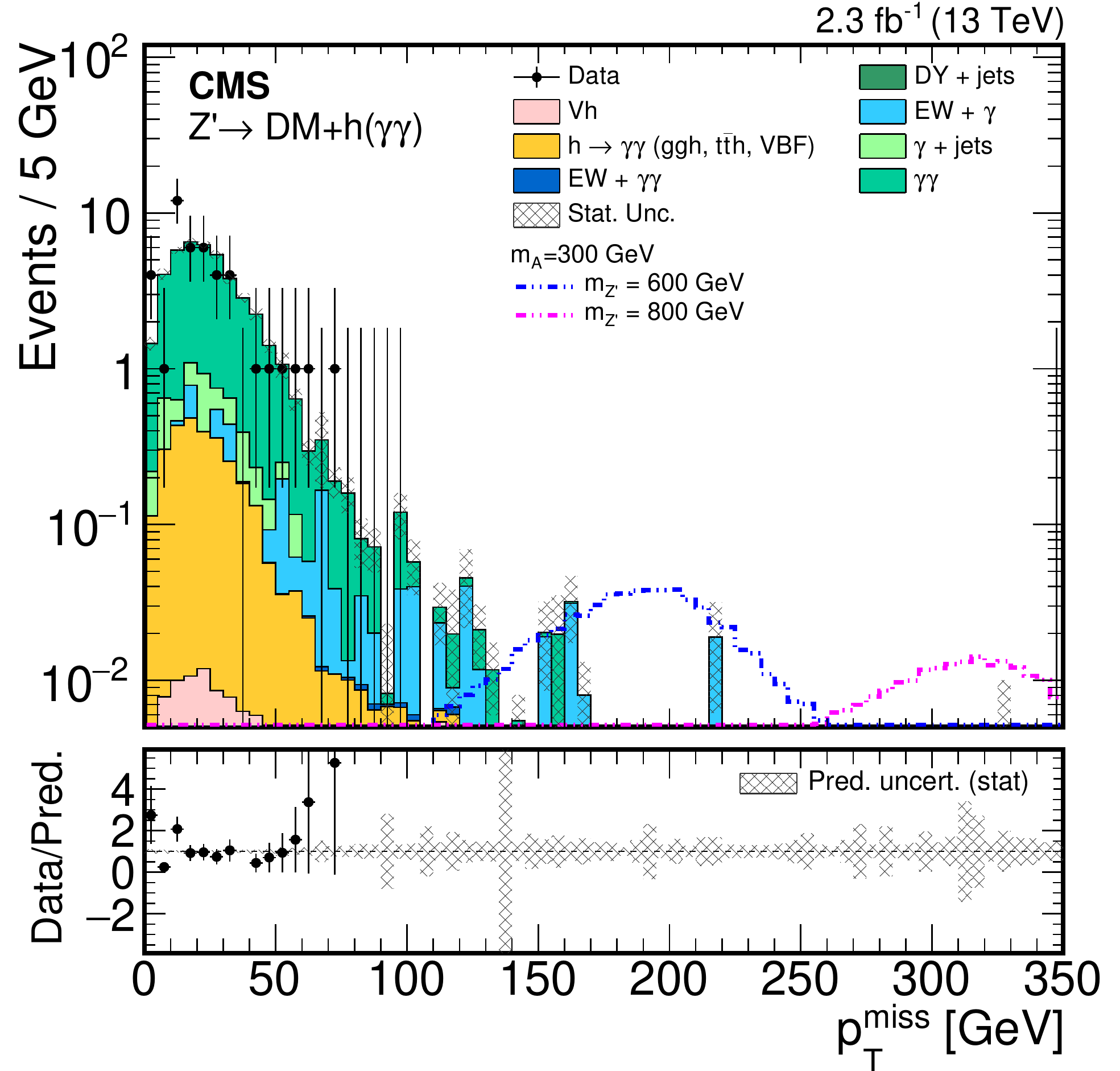}
\caption{\label{fig:mggmet_dist}
Expected and observed distribution of $m_{\gamma\gamma}$ (left) in events passing all selection criteria except the $m_{\gamma\gamma}$ and \MET  requirement.
Expected and observed distribution of \MET (right) for events passing all selection criteria including  $120\GeV <m_{\gamma\gamma}< 130\GeV$ except \MET requirement.
Two different \mzp signal points are overlaid.
Other parameters for this model are fixed to $m_{\chi} = 100\GeV$ and $\tan{\beta} = g_{\chi} = 1$. The cross sections
for the signal models are computed assuming $\gzp = 0.8$.
For both plots, the total simulated background is normalized to the total number of events in data.
The bottom panels show the data-to-simulation ratios for background predictions with a hatched band corresponding to the uncertainty due to the finite size of simulated samples.}

\end{figure}

The SR of this analysis is defined as the region with $120 < m_{\gamma\gamma} <130\GeV$ and \MET above 105\GeV.
The distribution of $m_{\gamma\gamma}$  for the selected events before the \MET requirement is shown in Fig.~\ref{fig:mggmet_dist} for the full mass range considered in this analysis: $105 < m_{\gamma\gamma} <  180\GeV$.
Also shown is the \MET distribution of the selected events after the $m_{\gamma\gamma}$ SR selection.
It can be seen that after applying the requirement that $m_{\gamma\gamma}$ has to be close to the Higgs boson mass, the SM background contribution in the high-\MET region is close to zero and the DM signal is well separated from the background distribution.

\subsubsection{Background estimation}
\label{sec:bkg_model}

The final state with a $\gamma\gamma$ pair and large \MET has two classes of background: resonant and
nonresonant. The contributions from each class are treated differently.

Resonant backgrounds arise from decays of the SM Higgs boson to two photons.
They appear as an additional peak under the expected signal peak and are evaluated with the MC simulation by counting the number of expected events from all SM Higgs production modes in the SR.

The contribution of the nonresonant backgrounds ($N^\text{bkg}_\mathrm{SB}$) in the sideband (SB) region,
mostly multijets and EW processes with mismeasured large \MET and misidentified photons,
is evaluated from the data by counting the number of events in the $m_{\gamma\gamma}$ sidebands $105 < m_{\gamma\gamma} <  120\GeV$ and
$130 < m_{\gamma\gamma} < 180\GeV$, with $ \MET > 105\GeV$ in both cases.
Then $N^\text{bkg}_\mathrm{SB}$ is scaled by a transfer factor $\alpha$ to take into account the relative
fraction between the number of events in the $m_{\gamma\gamma}$ SR and SB region.
The expected number of nonresonant background events in the SR is given by:
\begin{equation}
\label{eq:alpha}
N^\text{bkg}_\mathrm{SR} = \alpha N^\text{bkg}_\mathrm{SB}.
\end{equation}

\begin{figure}
\centering
\includegraphics[width=0.5\textwidth]{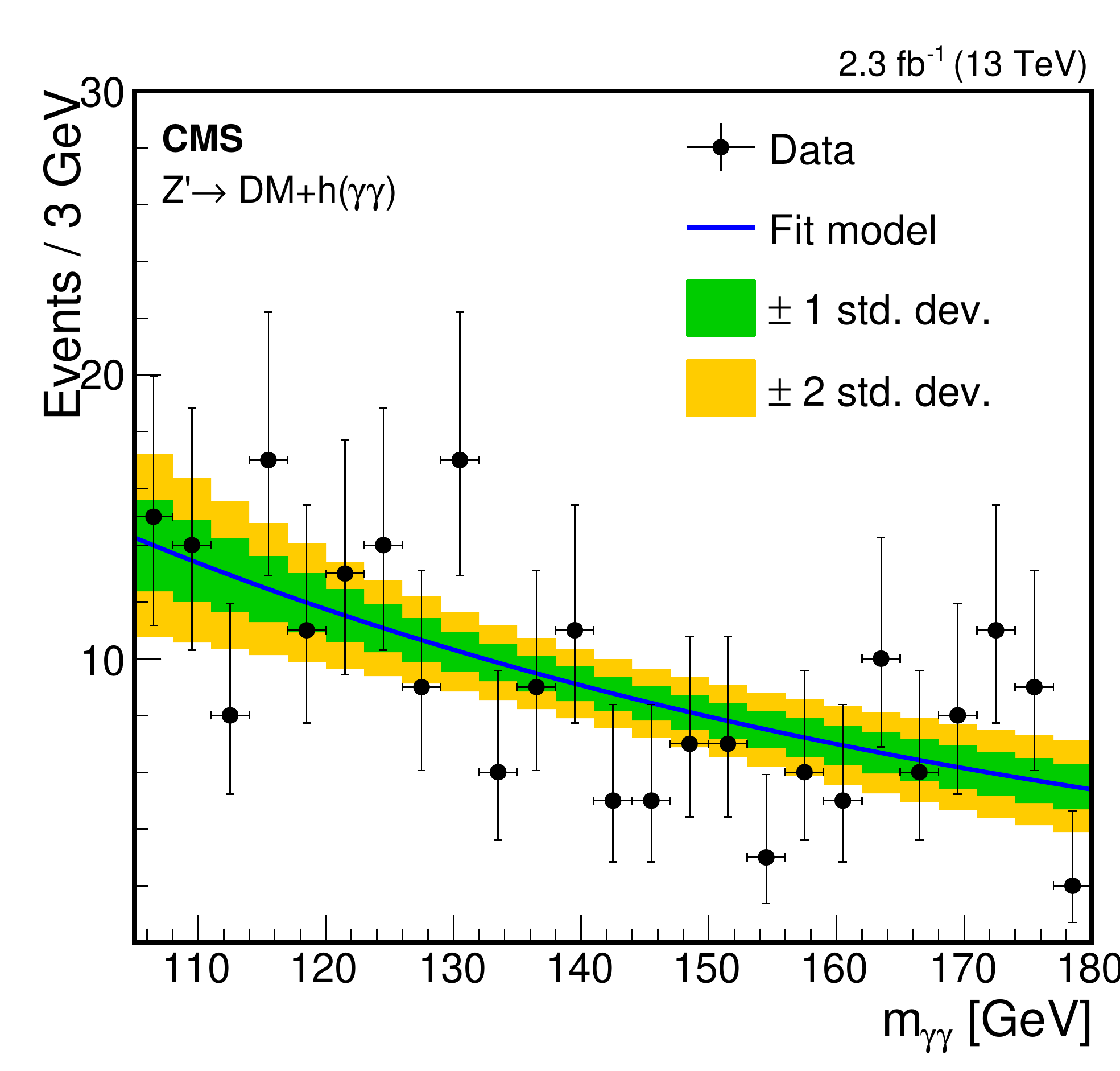}
\caption{\label{fig:fits}
Fit to the diphoton invariant mass distribution in the low-\MET CR in data used to evaluate $\alpha$. The function used is a power law with one free parameter.
The uncertainties in the background shapes associated with the statistical uncertainties of the
fit are shown by the one and two standard deviation bands.
}
\end{figure}

The derivation of $\alpha$
relies on the knowledge of the background shape~$f_\text{bkg}(m_{\gamma\gamma})$ as follows:
\begin{equation}
\label{eq:alphatwo}
\alpha = \frac{\int_\mathrm{SR}f_\text{bkg}(m_{\gamma\gamma})\rd m_{\gamma\gamma}}{\int_\mathrm{SB}f_\text{bkg}(m_{\gamma\gamma})\rd m_{\gamma\gamma}},
\end{equation}
and is evaluated by performing a fit to the $m_{\gamma\gamma}$ distribution in a CR of the data.
In this analysis, the low-\MET CR, with $ \MET < 105\GeV$, is used.
The fit to data in the low-\MET region used to calculate $\alpha$ is shown in Fig.~\ref{fig:fits}.
In this case the negligible contribution
of the resonant SM Higgs boson processes is not considered.
The data are fit with a background-only model using an analytic power law function:
\begin{equation}
f(x) = a x^{-b}
\label{eq:pl}
\end{equation}
where the parameter a, the normalization, and b are free parameters, defined as positive.
The fit is performed with an unbinned maximum likelihood technique.
The function defined in Eq.(\ref{eq:pl}) was chosen after examining several models and
performing a bias study using nonresonant background MC to evaluate any possible background mismodeling, following the procedure described in Ref.~\cite{highMassDiPhoton}.
It has been verified that the fitted parameters of the power law function are compatible within the uncertainties with both data and simulation.

To derive a robust estimate of $\alpha$, several fits to both data and simulated background
events are performed using different analytic functions and looking at different CRs of \MET.
Within the uncertainties, $\alpha$ is independent of the \MET CR used and is consistent between data and simulation.
The fitted shape of the low-\MET CR in data is taken as the nominal background shape.
This yields $\alpha = 0.190\pm 0.035$ (stat). Alternative analytic functions, as well as alternative \MET CRs in both data and simulation are considered in order to estimate the systematic uncertainty in this parameter, as described in Section~\ref{sec:systematics}.

\section{Systematic uncertainties}\label{sec:systematics}

The systematic uncertainties common to the two Higgs boson decay channels are as follows.

An uncertainty of 2.7\% is used for the normalization of simulated samples in order to reflect the uncertainty in the integrated luminosity measurement in 2015 \cite{CMS-PAS-LUM-15-001}.
In the \Hbb analysis an uncertainty of 2\% is estimated in the signal yield for \MET above 170\GeV by varying the parameters describing the trigger turn-on.
For the \HGG analysis the trigger uncertainty (approximately 1\%) is extracted from  $Z\to\Pep\Pem$ events using the tag-and-probe technique \cite{WZFirstPaper}.
The following uncertainties in clustered and unclustered calorimetric energy affect the \MET shapes and the normalization of the signal and background yield predictions:
the JES for each jet is varied within one standard deviation as a function of \PT and $\eta$, and the efficiency of the event selection is recomputed to assess the variation on the normalization and \MET shape for  signal and backgrounds;
the signal acceptance and efficiency are recomputed after smearing the energy of each jet to correct for the difference in jet energy resolution between the data and simulation (${\approx}5\%$);
the systematic uncertainties in the calibration of unclustered energy in the calorimeter are propagated as normalization and shape uncertainties in the \MET calculation.
The total effect of the systematic uncertainty in the signal yield, considering all of these variations on \MET is approximately 3\% for the \Hbb analysis and less than 1\% for the \HGG analysis. Among the three sources, the JES is the one that most affects the signal yield.

The following systematic uncertainties only affect the \Hbb decay channel:
The b tagging scale factors are applied consistently to jets in signal and background events. An average systematic uncertainty of 6\% per b jet, 12\% per c jet, and 15\%
per light quark or gluon jet is used to account for the normalization uncertainty \cite{BTV-15-001}.
The pruned mass distribution of the AK8 jet is not perfectly reproduced by simulation. Therefore, a control region, with a large number of events enriched in boosted hadronically decaying W bosons reconstructed as AK8 jets, is used to measure the systematic uncertainty due to this effect,
giving an estimated value of 5\%. Moreover, different hadronization algorithms (\PYTHIA and \HERWIGpp) give slightly different shapes for the pruned mass distribution. Therefore, an additional uncertainty of 10\% is assigned to account for the difference between simulations.
For the boosted regime, the same background normalization scale factor is used  for W+jets and top quark backgrounds. The uncertainty in the relative normalization of these two processes is 30\%.
An uncertainty of 2\% is measured by varying the lepton efficiency scale factors within one standard deviation and recomputing the signal selection efficiency.
For W+jets, Z($\to\nu\overline{\nu}$)+jets and top quark backgrounds, variations in the renormalization and factorization scales directly affect the normalization and shape of the \MET distribution. A variation of approximately 5\% is found for the yields of these backgrounds in the signal region.
The uncertainty in the signal acceptance and \MET shape due to the choice of PDFs is measured following the method described by the PDF4LHC group \cite{pdf4lhc}. A variation of approximately 3\% is found in the signal yields.
The effect of electroweak corrections as described in Section~\ref{sec:datasimulation} is studied by recomputing the normalization and
shapes for the W+jets and Z($\to \nu\overline{\nu}$)+jets backgrounds, by alternately removing the corrections or doubling them.
An uncertainty of 20\% is assumed for the single top quark , SM Higgs boson, and diboson production rates.
Uncertainties due to the finite size of the signal and background simulated samples
are included in the normalization and shape, such that each bin of the final fitted distributions is affected independently.

In summary, for \Hbb, the overall uncertainties related to background determination methods, simulation, and theory inputs are estimated to be 10\% in the background contributions in the SR.
The impact of the uncertainty in the major background contributions (W+jets, Z($\to \nu\overline{\nu}$)+jets and top quarks)
in the SR is reduced by constraining the normalizations of these processes in data with the simultaneous fit of \MET shapes in the SR and CRs.
The major sources of systematic uncertainties that affect the fit are JES uncertainties, b tagging uncertainties, and the statistical uncertainty in the
simulated Z($\to\nu\overline{\nu}$)+jets and W+jets background samples.
The effect of the remaining uncertainties  on the final fit is ${\approx}1\%$.

The following systematic uncertainties affect only the \HGG analysis:
\label{sec:rate_syst}
As shown in Equation~(\ref{eq:alpha}), the predicted number of nonresonant background events in the SR is evaluated from the number of observed events in the $m_{\gamma\gamma}$ sidebands in the high-\MET region ($N^\text{bkg}_\mathrm{SB}$) multiplied by a transfer factor $\alpha$ obtained by fitting the $m_{\gamma\gamma}$ distribution in the low-\MET control region. Therefore two different systematic uncertainties are assigned to this procedure, one for $N^\text{bkg}_\mathrm{SB}$ and one for $\alpha$.
The first systematic uncertainty takes into account the fact that $N^\text{bkg}_\mathrm{SB}$ is statistically limited.
Secondly, a 20\% systematic uncertainty is assigned to reflect the imperfect knowledge of the background
$m_{\gamma\gamma}$ shape in the low-\MET region, hence on the knowledge of the $\alpha$ factor.
This uncertainty is obtained by performing the fit to the $m_{\gamma\gamma}$ distribution using several analytic functions, using data rather than using simulated events,
and using other \MET CRs.
An observed peak above the diphoton continuum in the $m_{\gamma\gamma}$ distribution around the SM Higgs boson mass would have a SM \HGG contribution.
In order to extract the DM signal, the resonant background contribution has to be evaluated and subtracted.
The SM Higgs boson contribution is affected by
both theoretical and experimental systematic uncertainties.
For each SM Higgs boson production mechanism (ggh, VBF, tth, Vh), the uncertainties on the PDFs and $\alpha_s$, provided in Ref.~\cite{yellowReport}, are addressed using the procedure from the PDF4LHC group~\cite{pdf4lhc}.
The size of the systematic uncertainty is computed for each process and category separately by checking the effect of each weight on the final event yield.
An additional uncertainty on the \HGG branching fraction of 5\% is included
following Ref.~\cite{yellowReport}.
A 1\% photon energy scale uncertainty is assigned. This number takes into account the knowledge of the energy scale at the Z boson peak and of its extrapolation to higher masses.
The uncertainty on the photon resolution correction factors is evaluated by raising and lowering the estimated additional Gaussian smearing measured at the Z boson peak by 0.5\% in quadrature.
The photon identification uncertainty is taken as an uncertainty in the data-to-simulation scale factors, which can be as large as 2\%, depending on the \pt and the $\eta$ of the photon.

The \HGG decay channel results are only marginally affected by systematic uncertainties as statistical uncertainties dominate the analysis.

\section{Results \label{sec:results}}
For the event selection described in Section~\ref{sec:eventselection}, the predicted signal acceptances multiplied by the efficiencies ($A \epsilon$) are listed in Table~\ref{tab:AcceptanceHGG} for the two decay channels.

\begin{table}[bthp]
\centering
\renewcommand{\arraystretch}{1.1}
\topcaption{\label{tab:AcceptanceHGG}
The product of acceptance and efficiency (with statistical uncertainty) for signal in the SR, after full event selection for the \Hbb (upper) and
the \HGG (lower) decay channels. The systematic uncertainty for \Hbb (\HGG) is approximately 10\% (5\%). For \Hbb, the value shown here is either for the resolved regime or for the boosted regime, depending on which is used for the calculation of the limit on $\sigma$ ($\PZpr \to \Az \Ph \to \chi \overline{\chi} \Ph $), as shown in Fig. \ref{fig:limit2d} left. }
\resizebox{\textwidth}{!}{
\begin{tabular}{c|c|c|c|c|c|c}
\begin{tabular}{c} \maz \\ $[\GeVns{}]$ \end{tabular} & 300 & 400 & 500 & 600 & 700 & 800 \\ \hline
\begin{tabular}{c} \mzp \\ $[\GeVns{}]$ \end{tabular} & \multicolumn{6}{c}{\Hbb} \\ \hline
\x600  & 0.058  $ \pm $ 0.003  & 0.013 $ \pm $ 0.003  &  ---      &  ---  & ---     & ---                                                               \\
\x800  & 0.132  $ \pm $ 0.003  & 0.117 $ \pm $ 0.003  & 0.083 $ \pm $ 0.003  & 0.040 $ \pm $ 0.003  & ---                    & ---                  \\
1000 & 0.245  $ \pm $ 0.004  & 0.218 $ \pm $ 0.003  & 0.167 $ \pm $ 0.002  & 0.123 $ \pm $ 0.003  & 0.181 $ \pm $ 0.003  & 0.066$ \pm $ 0.003 \\
1200 & 0.282  $ \pm $ 0.003  & 0.272 $ \pm $ 0.004  & 0.262 $ \pm $ 0.003  & 0.238 $ \pm $ 0.004  & 0.195 $ \pm $ 0.003  & 0.126$ \pm $ 0.003 \\
1400 & 0.286  $ \pm $ 0.003  & 0.287 $ \pm $ 0.003  & 0.283 $ \pm $ 0.003  & 0.279 $ \pm $ 0.003  & 0.285 $ \pm $ 0.003  & 0.249$ \pm $ 0.003 \\
1700 & 0.280  $ \pm $ 0.003  & 0.284 $ \pm $ 0.003  & 0.283 $ \pm $ 0.003  & 0.284 $ \pm $ 0.003  & 0.285 $ \pm $ 0.004  & 0.284$ \pm $ 0.003 \\
2000 & 0.269  $ \pm $ 0.005  & 0.271 $ \pm $ 0.003  & 0.275 $ \pm $ 0.003  & 0.273 $ \pm $ 0.003  & 0.276 $ \pm $ 0.003  & 0.279$ \pm $ 0.004 \\
2500 & 0.248  $ \pm $ 0.003  & 0.246 $ \pm $ 0.003  & 0.250 $ \pm $ 0.004  & 0.251 $ \pm $ 0.003  & 0.255 $ \pm $ 0.003  & 0.256$ \pm $ 0.003 \\
\hline
\begin{tabular}{c} \mzp \\ $[\GeVns{}]$ \end{tabular}  & \multicolumn{6}{c}{\HGG} \\
\hline
600     & 0.317 $ \pm $ 0.004 & 0.212 $ \pm $ 0.003 & --- & --- & --- & ---                                                                 \\
800     & 0.399 $ \pm $ 0.004 & 0.386 $ \pm $ 0.003 & 0.348 $ \pm $ 0.003 &  0.280 $\pm$ 0.003  & --- & ---                                 \\
1000 	& 0.444 $ \pm $ 0.004 & 0.437 $ \pm $ 0.003 & 0.422 $ \pm $ 0.003 & 0.402 $ \pm $ 0.003 & 0.373 $ \pm $ 0.003 & 0.330 $ \pm $ 0.003 \\
1200 	& 0.474 $ \pm $ 0.004 & 0.468 $ \pm $ 0.003 & 0.461 $ \pm $ 0.003 & 0.454 $\pm$ 0.003  & 0.438 $ \pm $ 0.003 & 0.417 $ \pm $ 0.003  \\
1400 	& 0.492 $ \pm $ 0.004 & 0.493 $ \pm $ 0.003 & 0.485 $ \pm $ 0.003 & 0.481 $ \pm $ 0.003 & 0.472 $ \pm $ 0.003 & 0.465 $ \pm $ 0.003 \\
1700 	& 0.493 $ \pm $ 0.004 & 0.499 $ \pm $ 0.003 & 0.504 $ \pm $ 0.003 & 0.503 $ \pm $ 0.003 & 0.499 $ \pm $ 0.003 & 0.498 $ \pm $ 0.003 \\
2000 	& 0.351 $ \pm $ 0.004 & 0.373 $ \pm $ 0.003 & 0.394 $ \pm $ 0.003 & 0.421 $ \pm $ 0.003 & 0.453 $ \pm $ 0.003 & 0.488 $ \pm $ 0.003 \\
2500 	& 0.213 $ \pm $ 0.004 &  0.217 $\pm$ 0.003  & 0.227 $ \pm $ 0.003 & 0.236 $ \pm $ 0.003 & 0.254 $ \pm $ 0.003 & 0.268 $ \pm $ 0.003 \\
\end{tabular}
}
\end{table}

\begin{table}[http]
\topcaption{Post-fit background event yields and observed numbers of events in data for 2.3\fbinv in both the resolved and the boosted regimes for the \Hbb analysis. The expected numbers of signal events for $m_{\mathrm{A}} = 300\GeV$, scaled to the nominal cross section with $\gzp = 0.8$, are also reported. The statistical and systematic uncertainties are shown separately in that order.
\label{tab:eventYieldTable} }
\small
\begin{center}
\begin{tabular}{c|cc}
{\Hbb analysis}& \multicolumn{2}{c}{Number of events (in 2.3\fbinv)}\\
\hline
Process                           & Resolved  & Boosted \\
\hline
$Z(\to \nu\overline{\nu})$+jets   & 29.6 $\pm$ 2.7 $\pm$ 4.1\x   &     19.3 $\pm$ 0.8  $\pm$ 1.8\x  \\
top quark                                 & 7.3 $\pm$ 1.8 $\pm$ 1.0      &     8.2 $\pm$ 1.7   $\pm$ 1.6                  \\
W+jets                                    & 9.1 $\pm$ 1.6 $\pm$ 1.5      &     10.7 $\pm$ 1.6  $\pm$ 2.0\x  \\
Diboson                                   & 2.7 $\pm$ 0.5 $\pm$ 0.5      &     1.5 $\pm$ 0.3   $\pm$ 0.4                 \\
Vh                                        & 2.0 $\pm$ 0.02 $\pm$ 0.2     &     0.8 $\pm$ 0.05  $\pm$ 0.2                  \\
Multijet                                  & 0.01 $\pm$ 0.01 $\pm$ 0.20  &     0.02 $\pm$ 0.01 $\pm$ 0.01              \\
\hline
Total background             &50.7 $\pm$ 2.9 $\pm$   4.6 \x  &     40.5 $\pm$ 2.4 $\pm$ 3.1\x   \\
\hline
Data                       & 44        & 38     \\
\multicolumn{3}{c}{}        \\ [-1.0ex]
 \mzp $[\GeVns{}]$           &      \multicolumn{2}{c}{}        \\
\hline
\x600                        &  29.0 $\pm$ 0.4 $\pm$ 3.5\x     & --- \\
\x800                        &  40.4 $\pm$ 0.5 $\pm$ 3.8\x     & ---   \\
1000                       &  23.3 $\pm$ 0.3 $\pm$  2.5\x & ---  \\
1200                       & ---    & 23.6 $\pm$ 0.4  $\pm$  2.4\x  \\
1400                       & ---    & 13.1 $\pm$ 0.3  $\pm$   1.4\x  \\
1700                       & ---    & 5.6 $\pm$  0.2  $\pm$   0.7   \\
2000                       & ---    & 2.3 $\pm$  0.1   $\pm$  0.3   \\
2500                       & ---    & 0.24 $\pm$ 0.01  $\pm$  0.03  \\
\end{tabular}
\end{center}
\end{table}

Table \ref{tab:eventYieldTable} shows, for the \Hbb channel, the
SR post-fit yields for each background and signal mass point along with the sum of the statistical and systematic uncertainties for the resolved and boosted regimes. The total background uncertainty is approximately 10\% and mainly driven by the systematic uncertainty.

For the \HGG channel, when applying the event selection to the data, two events are observed in the $m_{\gamma\gamma}$ sidebands and are used to
evaluate the magnitude of the nonresonant background as described in Section~\ref{sec:bkg_model}.
This yields an expected number of $0.38 \pm 0.27\stat$ nonresonant background events in the SR.
Expected resonant background contributions are taken from the simulation as detailed in Section~\ref{sec:bkg_model} and are $0.057 \pm 0.006\stat$ events considering both the Vh production (dominant) and the gluon fusion mode. Zero events are observed in the SR in the data.

Since no excess of events has been observed over the SM background expectation in the signal region, the results of this search are interpreted in terms of an upper limit on the production 
of DM candidates in association with a Higgs boson 
in the process  $\PZpr \to \Az \Ph \to \chi \overline{\chi} \Ph$.
The upper limits are computed at 95\% confidence level (CL) using a modified frequentist method (CL$_s$) \cite{yellowReport, bib:CLS1, bib:CLS2} computed with an asymptotic approximation \cite{bib:CLS3}.
A profile likelihood ratio is used as the test statistic in which systematic uncertainties are modeled as nuisance parameters.
These limits are obtained as a function of \mzp and \maz for both Higgs boson decay channels and for the combination of the two.
The two decay channels are combined using the branching ratios predicted by the SM.
In the combination of the two analyses, all signal and \MET-related systematic uncertainties as well as the systematic uncertainty in the integrated luminosity
are assumed to be fully correlated.

\begin{figure}[htbp]
\centering
\includegraphics[width=0.45\textwidth]{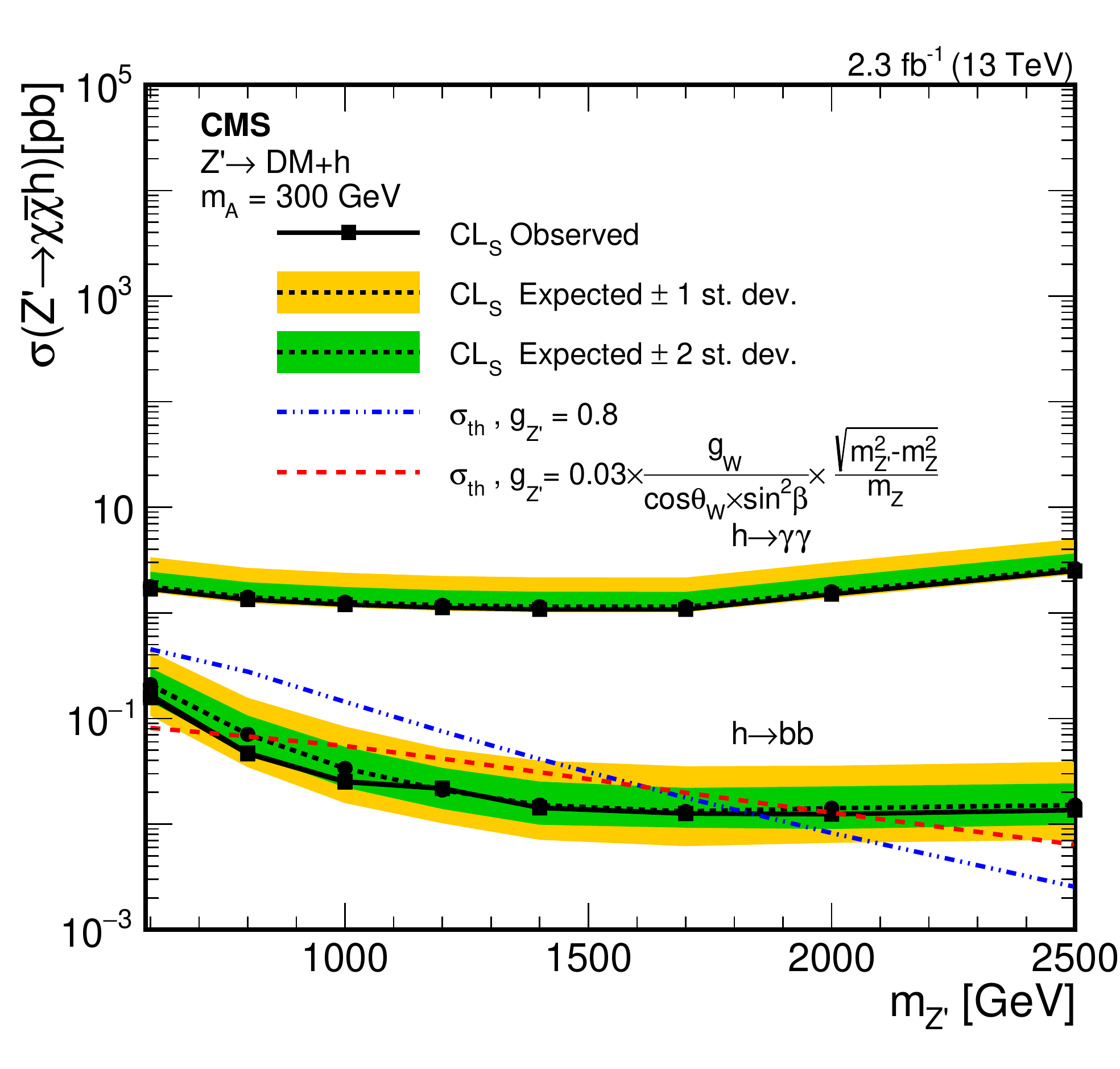}
\includegraphics[width=0.45\textwidth]{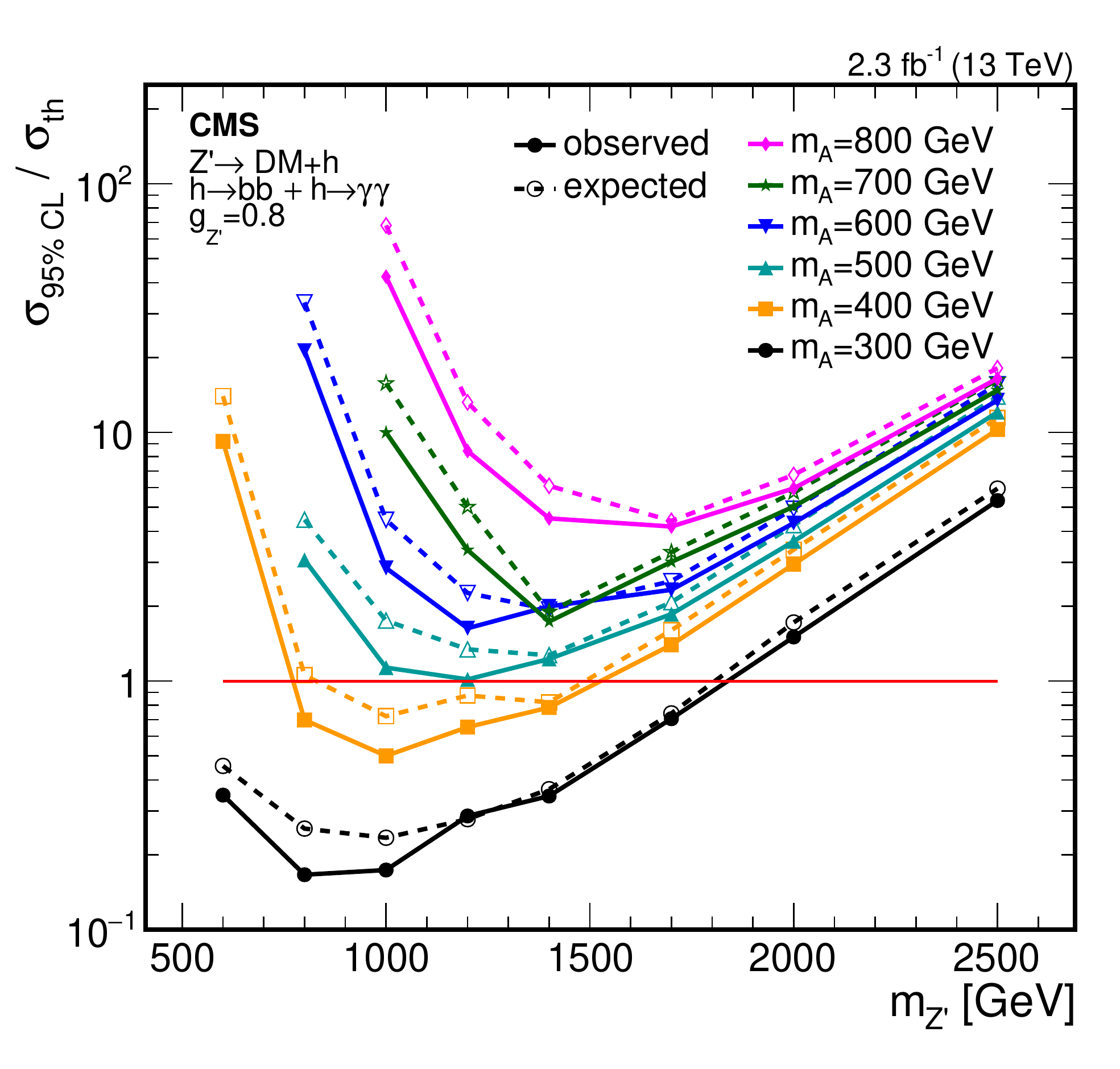}
\caption{Left: The expected and observed 95\% CL limits
on dark matter production cross sections for \Hbb and \HGG for \maz = 300\GeV. The exclusion region is shown for two
\gzp values. The dark green and light yellow  bands show the 68\% and 95\% uncertainties on the expected limit.
Right: The expected and observed 95\% CL limits on the signal strength for \maz = 300--800\GeV are shown.
Other parameters for this model are fixed to $m_{\chi} = 100\GeV$ and $\tan{\beta} = g_{\chi} = 1$.
The theoretical cross section ($\sigma_{\mathrm{th}}$) used for the right-hand plot is calculated using \gzp = 0.8.}
\label{fig:limitsexpected}
\end{figure}

Figure~\ref{fig:limitsexpected}  (left) shows the 95\% CL expected and observed limits on the dark matter production cross section $\sigma(\PZpr \to \Az \Ph \to \chi \overline{\chi} \Ph )$, for \Hbb and \HGG for \maz = 300\GeV. These results, obtained with $m_{\chi}= 100\GeV$, can be considered valid for any dark matter particle mass below 100\GeV since the branching fraction for decays of \Az to DM particles, $\mathcal{B}(\Az\to\chi\overline{\chi})$, decreases as $m_{\chi}$ increases.
As shown in Figure~\ref{fig:limitsexpected}, for the phase space parameters considered for this model ($g_{\chi}$ and $\tan{\beta}$ equal to unity), results of the combined analysis are mainly driven by the \Hbb channel. The combination with the \HGG channel provides a 2-4\% improvement in terms of constraints on the model for the low \PZpr mass values. Future iterations of this search will explore additional phase space regions of the \cPZpr-2HDM model, i.e. larger values of $\tan{\beta}$, where the \HGG channel becomes more sensitive than \Hbb~\cite{2HDM}.

\begin{figure}[htbp]
\centering
\includegraphics[width=0.45\textwidth]{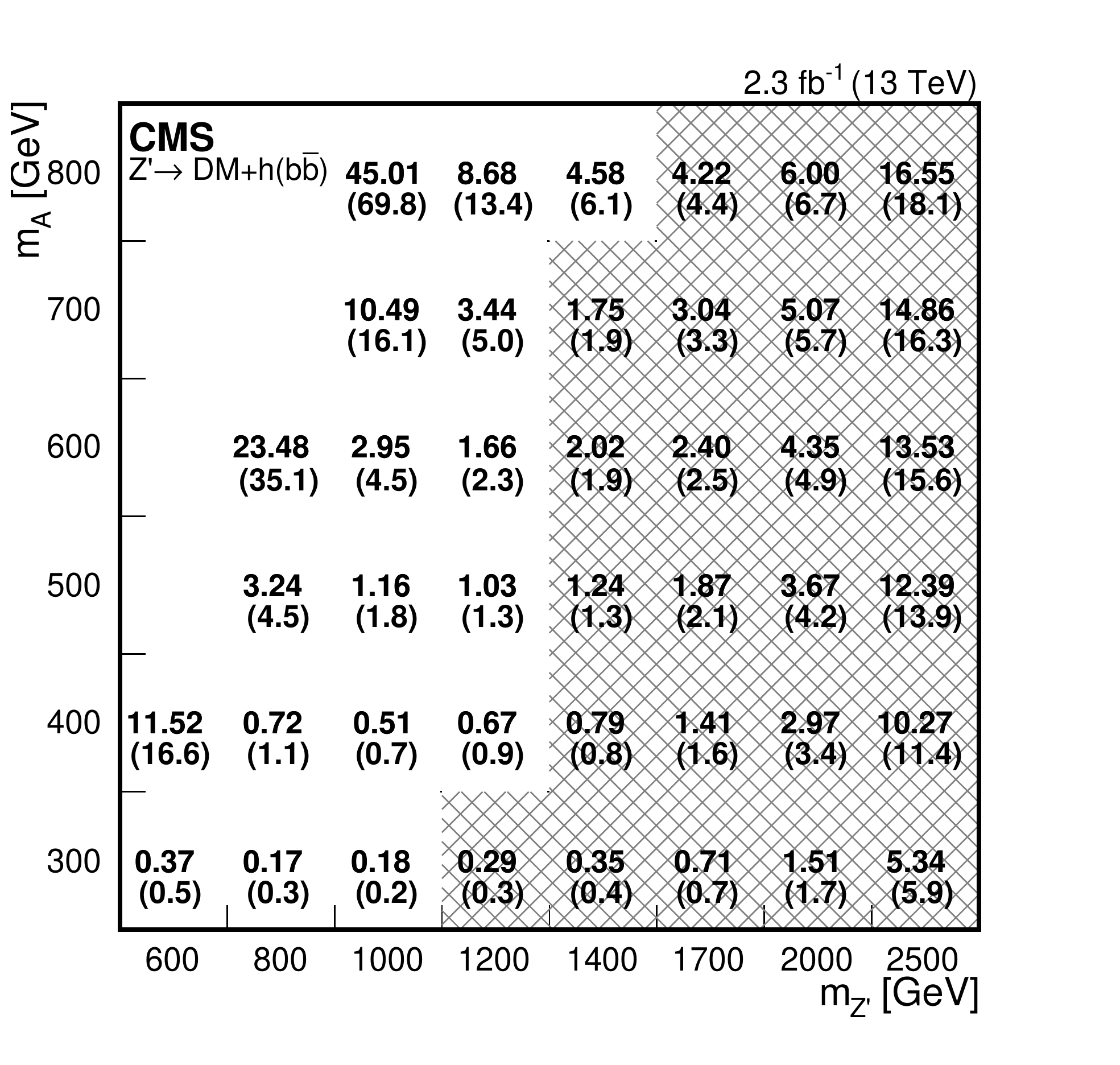}
\includegraphics[width=0.45\textwidth]{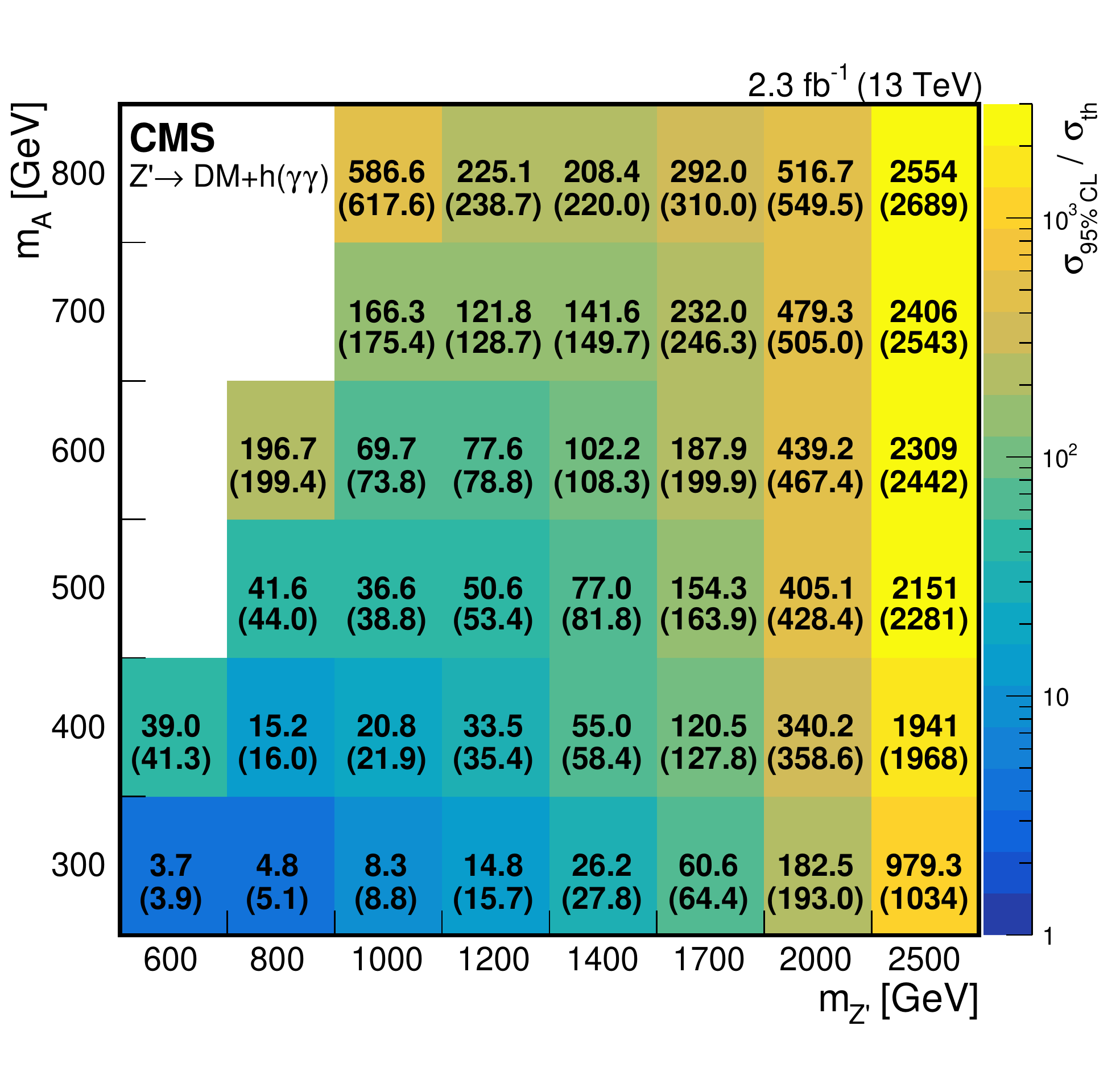}
\caption{The observed (expected) 95\% CL limits on the signal strength (as in Fig.~\ref{fig:limitsexpected} right), separately for the \Hbb (left) and \HGG (right) decay channels, and for \maz = 300--800\GeV and \mzp = 600--2500\GeV. Other parameters for this model are fixed to $m_{\chi} = 100\GeV$ and $\tan{\beta} = g_{\chi} = 1$. The theoretical cross sections
are calculated using $g_{\zp} = 0.8$. For \Hbb, the results for the resolved analysis are shown over a white background, whereas the boosted analysis results are shown over a hatched background. }
\label{fig:limit2d}
\end{figure}

\begin{figure}[htbp]
\centering
\includegraphics[width=0.45\textwidth]{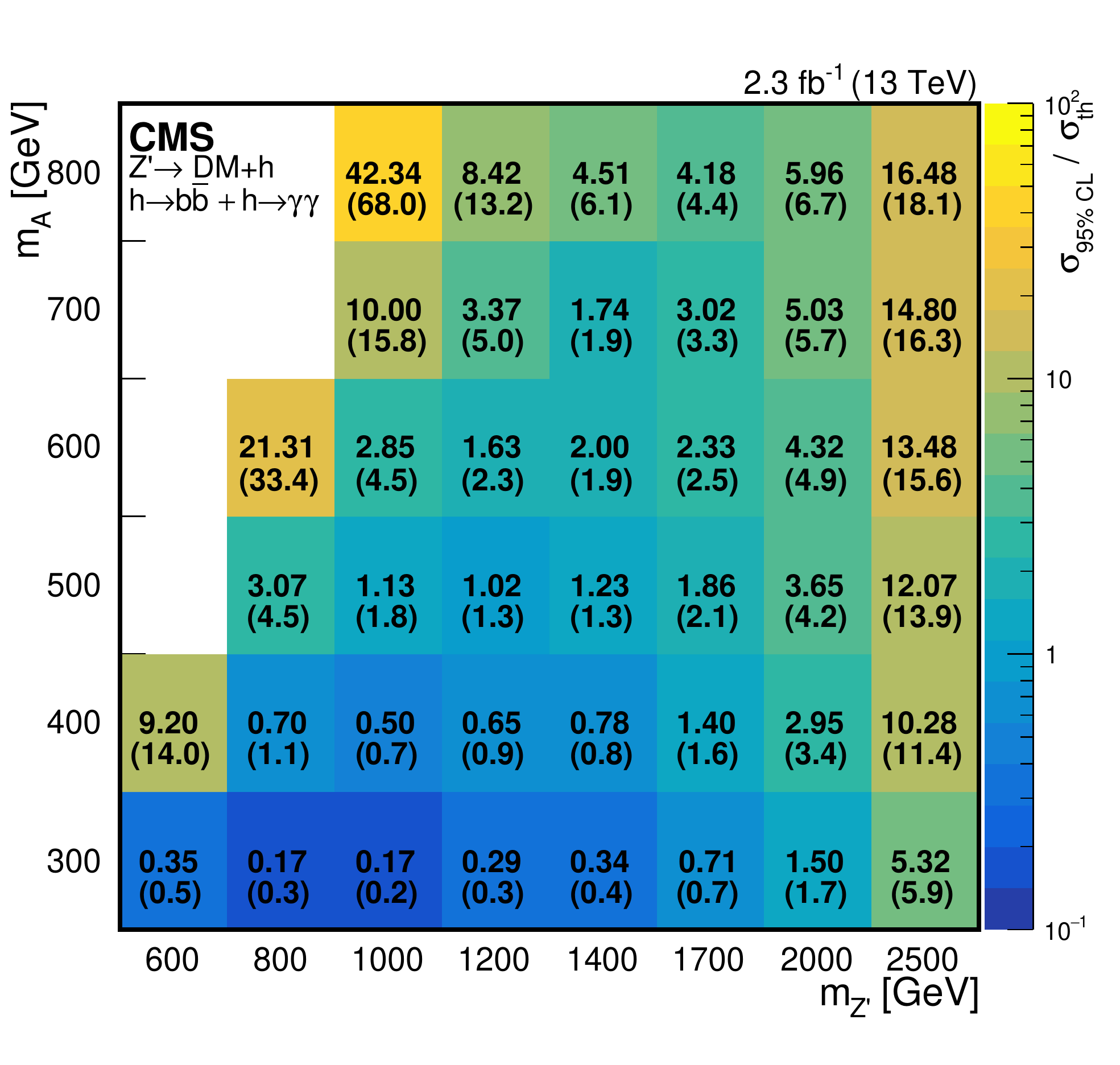}
\caption{The observed (expected) 95\% CL limits on the signal strength (as in Fig.~\ref{fig:limitsexpected} right) for the combination of \HGG and \Hbb decay channels, and for \maz = 300--800\GeV and \mzp = 600--2500\GeV. Other parameters for this model are fixed to $m_{\chi} = 100$ \GeV and $\tan{\beta} = g_{\chi} = 1$. The theoretical cross sections times branching fractions are calculated using $g_{\zp} = 0.8$.}
\label{fig:limit2dcombo}
\end{figure}

Figures~\ref{fig:limitsexpected} (right) and~\ref{fig:limit2d} show the 95\% CL  expected and observed upper limits on the signal strength $\sigma_{95\% \mathrm{CL}} (\PZpr \to \Az \Ph \to \chi \overline{\chi} \Ph )/\sigma_\text{theory} (\PZpr \to \Az \Ph \to \chi \overline{\chi} \Ph )$.
For \maz = 300\GeV, the \PZpr mass range from 600 to 1780\GeV is expected to be excluded with a 95\% CL when the signal model
cross section is calculated using \gzp = 0.8, while the
observed data, for \maz = 300\GeV, exclude the \zp mass range from 600 to 1860\GeV.
When the signal model cross section is calculated using the constrained \gzp, the expected exclusion range is 830 to 1890\GeV,
and the observed exclusion range is 770 to 2040\GeV.
Figure~\ref{fig:limit2d} shows the expected and observed upper limits on the signal strength for the \Hbb and \HGG decay channels.
Figure~\ref{fig:limit2dcombo} shows the upper limits on the signal strength combining the results from both the \Hbb and \HGG decay channels.

\section{Summary}\label{sec:conclusion}

A search has been performed for dark matter produced in association with a Higgs boson.
The analysis is based on 2.3\fbinv of proton-proton collision data collected by the CMS
experiment at $\sqrt{s} = 13\TeV$.
This analysis focuses on a \cPZpr-2HDM model in which the \cPZpr decays to a light SM-like scalar Higgs boson and
a pseudoscalar boson \Az, that in turn decays to two dark matter candidates.
Two distinct channels are studied, where the Higgs boson decays to two b quarks or two photons.

No significant deviation is observed from the standard model background.  With optimized selections, limits on the signal cross section $\sigma(\PZpr \to \Az \Ph \to \chi \overline{\chi} \Ph)$ are calculated for various values of \mzp and \maz assuming $g_{\chi}$ and $\tan{\beta}$ equal to one. The limits are valid for any dark matter particle mass below 100\GeV.
For $\maz = 300\GeV$, the observed data exclude the \zp mass range of 600 to 1860\GeV for $\gzp = 0.8$, and the range 770 to 2040\GeV for the constrained value of \gzp. This is the first result on a search for dark matter produced in association with a Higgs boson at $\sqrt{s}$ = 13\TeV that combines results from the \Hbb and \HGG channels.

\begin{acknowledgments}

\hyphenation{Bundes-ministerium Forschungs-gemeinschaft Forschungs-zentren Rachada-pisek} We congratulate our colleagues in the CERN accelerator departments for the excellent performance of the LHC and thank the technical and administrative staffs at CERN and at other CMS institutes for their contributions to the success of the CMS effort. In addition, we gratefully acknowledge the computing centres and personnel of the Worldwide LHC Computing Grid for delivering so effectively the computing infrastructure essential to our analyses. Finally, we acknowledge the enduring support for the construction and operation of the LHC and the CMS detector provided by the following funding agencies: the Austrian Federal Ministry of Science, Research and Economy and the Austrian Science Fund; the Belgian Fonds de la Recherche Scientifique, and Fonds voor Wetenschappelijk Onderzoek; the Brazilian Funding Agencies (CNPq, CAPES, FAPERJ, and FAPESP); the Bulgarian Ministry of Education and Science; CERN; the Chinese Academy of Sciences, Ministry of Science and Technology, and National Natural Science Foundation of China; the Colombian Funding Agency (COLCIENCIAS); the Croatian Ministry of Science, Education and Sport, and the Croatian Science Foundation; the Research Promotion Foundation, Cyprus; the Secretariat for Higher Education, Science, Technology and Innovation, Ecuador; the Ministry of Education and Research, Estonian Research Council via IUT23-4 and IUT23-6 and European Regional Development Fund, Estonia; the Academy of Finland, Finnish Ministry of Education and Culture, and Helsinki Institute of Physics; the Institut National de Physique Nucl\'eaire et de Physique des Particules~/~CNRS, and Commissariat \`a l'\'Energie Atomique et aux \'Energies Alternatives~/~CEA, France; the Bundesministerium f\"ur Bildung und Forschung, Deutsche Forschungsgemeinschaft, and Helmholtz-Gemeinschaft Deutscher Forschungszentren, Germany; the General Secretariat for Research and Technology, Greece; the National Scientific Research Foundation, and National Innovation Office, Hungary; the Department of Atomic Energy and the Department of Science and Technology, India; the Institute for Studies in Theoretical Physics and Mathematics, Iran; the Science Foundation, Ireland; the Istituto Nazionale di Fisica Nucleare, Italy; the Ministry of Science, ICT and Future Planning, and National Research Foundation (NRF), Republic of Korea; the Lithuanian Academy of Sciences; the Ministry of Education, and University of Malaya (Malaysia); the Mexican Funding Agencies (BUAP, CINVESTAV, CONACYT, LNS, SEP, and UASLP-FAI); the Ministry of Business, Innovation and Employment, New Zealand; the Pakistan Atomic Energy Commission; the Ministry of Science and Higher Education and the National Science Centre, Poland; the Funda\c{c}\~ao para a Ci\^encia e a Tecnologia, Portugal; JINR, Dubna; the Ministry of Education and Science of the Russian Federation, the Federal Agency of Atomic Energy of the Russian Federation, Russian Academy of Sciences, the Russian Foundation for Basic Research and the Russian Competitiveness Program of NRNU ``MEPhI"; the Ministry of Education, Science and Technological Development of Serbia; the Secretar\'{\i}a de Estado de Investigaci\'on, Desarrollo e Innovaci\'on, Programa Consolider-Ingenio 2010, Plan de Ciencia, Tecnolog\'{i}a e Innovaci\'on 2013-2017 del Principado de Asturias and Fondo Europeo de Desarrollo Regional, Spain; the Swiss Funding Agencies (ETH Board, ETH Zurich, PSI, SNF, UniZH, Canton Zurich, and SER); the Ministry of Science and Technology, Taipei; the Thailand Center of Excellence in Physics, the Institute for the Promotion of Teaching Science and Technology of Thailand, Special Task Force for Activating Research and the National Science and Technology Development Agency of Thailand; the Scientific and Technical Research Council of Turkey, and Turkish Atomic Energy Authority; the National Academy of Sciences of Ukraine, and State Fund for Fundamental Researches, Ukraine; the Science and Technology Facilities Council, UK; the US Department of Energy, and the US National Science Foundation.

Individuals have received support from the Marie-Curie programme and the European Research Council and EPLANET (European Union); the Leventis Foundation; the A. P. Sloan Foundation; the Alexander von Humboldt Foundation; the Belgian Federal Science Policy Office; the Fonds pour la Formation \`a la Recherche dans l'Industrie et dans l'Agriculture (FRIA-Belgium); the Agentschap voor Innovatie door Wetenschap en Technologie (IWT-Belgium); the Ministry of Education, Youth and Sports (MEYS) of the Czech Republic; the Council of Scientific and Industrial Research, India; the HOMING PLUS programme of the Foundation for Polish Science, cofinanced from European Union, Regional Development Fund, the Mobility Plus programme of the Ministry of Science and Higher Education, the National Science Center (Poland), contracts Harmonia 2014/14/M/ST2/00428, Opus 2014/13/B/ST2/02543, 2014/15/B/ST2/03998, and 2015/19/B/ST2/02861, Sonata-bis 2012/07/E/ST2/01406; the National Priorities Research Program by Qatar National Research Fund; the Programa Clar\'in-COFUND del Principado de Asturias; the Thalis and Aristeia programmes cofinanced by EU-ESF and the Greek NSRF; the Rachadapisek Sompot Fund for Postdoctoral Fellowship, Chulalongkorn University and the Chulalongkorn Academic into Its 2nd Century Project Advancement Project (Thailand); and the Welch Foundation, contract C-1845.

\end{acknowledgments}

\clearpage

\bibliography{auto_generated}

\cleardoublepage \appendix\section{The CMS Collaboration \label{app:collab}}\begin{sloppypar}\hyphenpenalty=5000\widowpenalty=500\clubpenalty=5000\input{EXO-16-012-authorlist.tex}\end{sloppypar}
\end{document}

%% file: EXO-16-012-authorlist.tex
\textbf{Yerevan Physics Institute,  Yerevan,  Armenia}\\*[0pt]
A.M.~Sirunyan, A.~Tumasyan
\vskip\cmsinstskip
\textbf{Institut f\"{u}r Hochenergiephysik,  Wien,  Austria}\\*[0pt]
W.~Adam, E.~Asilar, T.~Bergauer, J.~Brandstetter, E.~Brondolin, M.~Dragicevic, J.~Er\"{o}, M.~Flechl, M.~Friedl, R.~Fr\"{u}hwirth\cmsAuthorMark{1}, V.M.~Ghete, C.~Hartl, N.~H\"{o}rmann, J.~Hrubec, M.~Jeitler\cmsAuthorMark{1}, A.~K\"{o}nig, I.~Kr\"{a}tschmer, D.~Liko, T.~Matsushita, I.~Mikulec, D.~Rabady, N.~Rad, B.~Rahbaran, H.~Rohringer, J.~Schieck\cmsAuthorMark{1}, J.~Strauss, W.~Waltenberger, C.-E.~Wulz\cmsAuthorMark{1}
% ...
\vskip\cmsinstskip
% ...
1:~~Also at Vienna University of Technology, Vienna, Austria\\
% ...
